\documentclass[fleqn,usenatbib]{mnras}

\usepackage[T1]{fontenc}
\usepackage{ae,aecompl}

\usepackage[utf8]{inputenc}

\usepackage{graphicx}	

\usepackage{bm}
\usepackage{soul}
\usepackage{amsfonts}
\usepackage{amssymb}
\usepackage{newtxtext,newtxmath}

\usepackage{hyperref} 
\usepackage{amsmath}
\usepackage{mathtools}
\usepackage{color}
\usepackage[nottoc,notlot,notlof]{tocbibind}
\DeclareMathOperator{\argmin}{\mathop{argmin}_{\alpha}}

\DeclarePairedDelimiter{\norm}{\lVert}{\rVert} 
\usepackage[center]{caption}
\usepackage{pdfpages}

\def\apj{Astrophys. J.}

\def\apjl{Astrophys. J. Lett.}
\def\apjs{Astrophys. J. Suppl. Ser.} 
\def\aap{Astron. Astrophys.}
\def\mnras{Mon. Not. R. Astron. Soc.} 
\def\prd{Phys. Rev. D}
\def\prl{Phys. Rev. Lett.}

\title[Classification of the core-collapse supernova explosion mechanism with learned dictionaries]{Classification of the core-collapse supernova explosion mechanism with learned dictionaries}

\author[Ainara Saiz-P\'erez, Alejandro Torres-Forn\'e, and Jos\'e A.~Font]
{Ainara Saiz-P\'erez$^{1}$,
Alejandro Torres-Forn\'e$^{1}$\thanks{E-mail: alejandro.torres@uv.es},
and Jos\'e A. Font$^{1,2}$
\\
$^{1}$Departamento de Astronom\'ia y Astrof\'isica, Universitat de Val\`encia, Dr. Moliner, 50, Burjassot (Val\`encia) E46100, Spain. \\
$^{2}$Observatori Astron\`omic, Universitat de Val\`encia, Catedr\'atico Jos\'e Beltr\'an 2, 46980, Paterna (Val\`encia) E46100, Spain. \\
}

\date{Accepted XXX. Received YYY; in original form ZZZ}

\pubyear{2021}

\begin{document}
\label{firstpage}
\pagerange{\pageref{firstpage}--\pageref{lastpage}}
\maketitle

\begin{abstract}
Core-collapse supernovae (CCSN) are a prime source of gravitational waves. Estimations of their typical frequencies make them perfect targets for the current network of advanced, ground-based detectors.
A successful detection could potentially reveal the underlying explosion mechanism through the analysis of the waveform. This has been illustrated using the Supernova Model Evidence Extractor (SMEE;~\citet{Logue:2012}), an algorithm based on principal component analysis and Bayesian model selection. Here, we present a complementary approach to SMEE based on (supervised) dictionary-learning and show that it is able to reconstruct and classify CCSN signals according to their morphology. Our waveform signals are obtained from (a) two publicly available catalogs built from numerical simulations of neutrino-driven ({\tt Mur}) and magneto-rotational ({\tt Dim}) CCSN explosions and (b) from a third `mock' catalog of simulated sine-Gaussian ({\tt SG}) waveforms. Those signals are injected into coloured Gaussian noise to simulate the background noise of Advanced LIGO in its broadband configuration and scaled to a freely-specifiable signal-to-noise ratio (SNR). We show that our approach correctly classifies signals from all three dictionaries. In particular, for SNR=15-20, we obtain perfect matches for both {\tt Dim} and {\tt SG} signals and about 85\% true classifications for {\tt Mur} signals. These results are comparable to those reported by SMEE for the same CCSN signals when those are injected in only one LIGO detector. We discuss the main limitations of our approach as well as possible improvements.
\end{abstract}



\section{Introduction}

The detection of the gravitational-wave (GW) signal GW150914 by Advanced LIGO~\citep{Aasi:2015} on September 14, 2015~\citep{abbott2016observation} inaugurated the field of GW astronomy. The observed waveform is in agreement with the signals predicted by numerical-relativity simulations of binary black hole (BBH) mergers. Two years later, the signal from a binary neutron star (BNS) merger, GW170817, was jointly detected by Advanced LIGO and Advanced Virgo~\citep{Acernese:2015,abbott2017gw170817}, an event accompanied by electromagnetic radiation famously observed by dozens of telescopes worldwide~\citep{MMA}. The LIGO-Virgo-KAGRA (LVK) Collaboration has already released three catalogs of GW transients comprising over 50 compact binary coalescence (CBC) confirmed detections~\citep{GWTC-1,GWTC-2,GWTC-2.1}. GW astronomy is allowing for an entirely new way to observe the Universe, one which could give us insight into astronomical phenomena which are not yet fully understood. Among these, the detection of GW signals emitted by core-collapse supernovae (CCSN) is highly anticipated. This is the type of GW source we discuss in this paper. 

GW data analysis has played a major role in the detection of all CBC events observed so far. The inspiral phase of quasi-circular compact binary mergers can be accurately described with generic waveform models derived from general relativity. On the contrary, CCSN waveforms are insufficiently modeled, which renders the strategy for their analysis entirely different to that of CBC sources. Searches of unmodeled sources are based on coherent time-frequency analysis of the data over a network of detectors, especially for long signals lasting several seconds. Moreover, unlike for CBC sources, it is currently not possible to relate the properties of CCSN progenitors (e.g.~mass, rotation rate, metallicity, magnetic fields, etc) with numerically-generated waveforms due to the large parameter space and uncertainties involved. In addition, owing to their intrinsic short duration, CCSN burst signals are likely to be affected by detector noise transients or `glitches'. Specific pipelines based on Bayesian inference have been developed by the LVK Collaboration to differentiate between signals and noise transients (see e.g.~\cite{cornish2015bayeswave,Kanner:2016,Littenberg:2016,Lynch:2016,Pannarale:2019,drago2021coherent}).  

Notwithstanding the intrinsic difficulties involved in the numerical modeling of CCSN, steady progress on waveform generation has been accomplished. The early bounce signal is the part of the waveform best understood since the early works of~\citet{Zwerger:1997} and~\citet{Dimmelmeier:2002}. The frequency of this part of the signal ($\sim 800$ Hz) is directly related to the rotational properties of the core~\citep{Dimmelmeier:2008,Abdikamalov:2014,Richers:2017,Takiwaki:2018,Pajkos:2019}. However, fast-rotating progenitors are not generic in all CCSN~\citep{Gossan:2016} and, moreover, the bounce signal is probably difficult to observe by current detectors due to its high frequency and low amplitude. However, the GW signal associated with the post-bounce evolution of the proto-neutron star (PNS) is more promising. This signal is sourced by the dynamical trigger of convection and oscillations due to downdrafts of accreting material and by the standing accretion shock instability (SASI)~\citep{Murphy:2009, Mueller:2013, Cerda-Duran:2013, Kuroda:2016, Andresen:2017,Takiwaki:2018,Andresen:2019,Radice:2019,Powell:2019,Shibagaki:2020,Powell:2020}. Waveforms typically last for $\sim 500$ ms until the onset of explosion; they may even extend beyond one second in the event of black hole formation. GW frequencies rise monotonically with time due to the contraction of the PNS. They can be as low as $\sim 100$ Hz, which turns them into a perfect target for advanced ground-based detectors. Recent progress on how to infer PNS properties from the analysis of GW data in CCSN are reported in~\cite{Torres-Forne:2019,Bizouard:2021}.

The LVK Collaboration has recently conducted an optically-targeted search of GW from CCSN within a distance of $\sim20$ Mpc during the first and second observing runs of Advanced LIGO and Advanced Virgo~\citep{abbott2020optically}. No significant GW candidate was detected in this search. Even a strong galactic CCSN, at a distance of $\sim10$ kpc, would likely result in a (dimensionless) GW strain of $\sim10^{-23}$~\citep{evans2017detecting}. For reference, GW150914 had a peak strain of $\sim10^{-21}$~\citep{abbott2016observation}. Outside of our Galaxy, certain CCSN models suggest signals might be observed from sources up to the distance of the Large Magellanic Cloud~\citep{abbott2020optically}.

Because the supernova explosion starts in the deepest parts of the star opaque to electromagnetic radiation, only GWs and neutrino emission can help understand its dynamics. Indeed, a neutrino burst was detected in association with supernova 1987A, in remarkable agreement with CCSN models~\citep{hirata1987observation}. This makes CCSN good targets for multimessenger astronomy, by combining the search for GWs with other types of messengers such as neutrino emission or electromagnetic signals (see~\citet{halim2021multimessenger} and references therein). Presently, two main possibilities are considered to explain the explosion, namely the neutrino-driven mechanism and the the magneto-rotational mechanism. Both lead to distinct features in the GW signal. As a result, a successful detection could potentially reveal the underlying explosion mechanism through the analysis of the waveform. This would also open the possibility to infer properties of the progenitor stars as well as of the PNS. Attempts to work out the feasibility of this proposal have already been reported. \citet{Logue:2012} performed a study based on principal component analysis (PCA) and Bayesian model selection over CCSN numerical signals and built an analysis algorithm called the Supernova
Model Evidence Extractor (SMEE) (see also~\citet{Brady:2004,Summerscales:2008,Heng:2009,Roever:2009} for earlier investigations). \citet{Logue:2012} found that the method was able to associate a potential GW signal injected into Gaussian noise with the magneto-rotational mechanism up to a distance of $\sim10$ kpc and with the neutrino mechanism up to only $\sim2$ kpc, using a single detector. More recently~\cite{Powell:2016} employed SMEE with CCSN signals in a three-detector network. The signals were injected into real noise from the initial Advanced LIGO and Advanced Virgo with their sensitivities altered to simulate the nominal design sensitivity of those detectors. They found that it is possible to determine if the explosion is driven by neutrino convection for sources within the Galaxy and driven by rapidly-rotating core collapse for sources up to to the Large Magellanic Cloud. Finally, \cite{Powell:2017} and~\citet{Powell:2018} further developed SMEE to incorporate the possibility of determining whether a candidate CCSN signal is a glitch before identifying the explosion mechanism, and \citet{roma2019astrophysics} extended SMEE to identify specific features in the GW signals associated with g-modes of oscillation of the PNS and SASI as well as using spectrograms of supernova waveforms rather than time-series waveforms.

Those past studies have motivated the present work. Here, we investigate the performance of an alternative approach to SMEE for the classification of the CCSN explosion mechanism through a (supervised) machine-learning algorithm based on learned dictionaries. We have recently applied this approach for reconstruction and classification of simulated GW signals and actual detector glitches, obtaining promising results~\citep{Torres:2016,llorens:2019,Torres-Forne:2020}. Our goal in this paper is to develop a method to denoise numerically-generated CCSN GW signals and to classify them depending on their morphology. To do so we use publicly available GW catalogs of simulated CCSN to train dictionaries and use them to denoise and reconstruct signals injected in Gaussian noise. For the sake of comparison with the results of~\cite{Logue:2012} we employ the same waveform catalogs. We also develop a method of classification to associate a given signal to a certain type of explosion mechanism and check the dependence of the results on the signal-to-noise ratio. On the whole our results are comparable to those obtained with SMEE when the same CCSN signals are injected in only one LIGO detector, as was done in~\cite{Logue:2012}.

This paper is organized as follows: Section 2 presents a brief description of the two prevailing CCSN explosion mechanisms along with an overview of the types of waveform signals we use in our study. In Section 3 we discuss the mathematical framework of our approach, providing details on dictionary-learning, signal denoising and classification. Section 4 deals with the process of training the dictionaries and selecting the optimal parameters of the method. Specific technical details are reported on Appendix A. In Section 5 we show the main results of our study. Finally, Section 6 contains our conclusions and outlines possible extensions of this work.

\section{Gravitational waves from CCSN}

CCSN are a prime source of GWs (see~\cite{Abdikamalov:2020} and references therein). Their complex dynamics are characterized by the existence of a time-dependent, mass-quadrupole moment which yields a non-zero GW strain. Such a time-dependent quadrupole originates from a number of processes, namely convection in the PNS and in the neutrino-heated hot bubble, anisotropic neutrino emission, and non-radial instabilities (e.g.~SASI). All of those are generic to all CCSN while additional mechanisms such as rapid rotation and magnetic fields are less generic. We next describe the two explosion mechanisms that are believed to operate in most CCSN, the neutrino mechanism and the magnetorotational mechanism, and discuss the two waveform catalogs we consider in our study.

\subsection{Neutrino-driven mechanism}

\begin{figure*}
\centering
\includegraphics[scale=0.35]{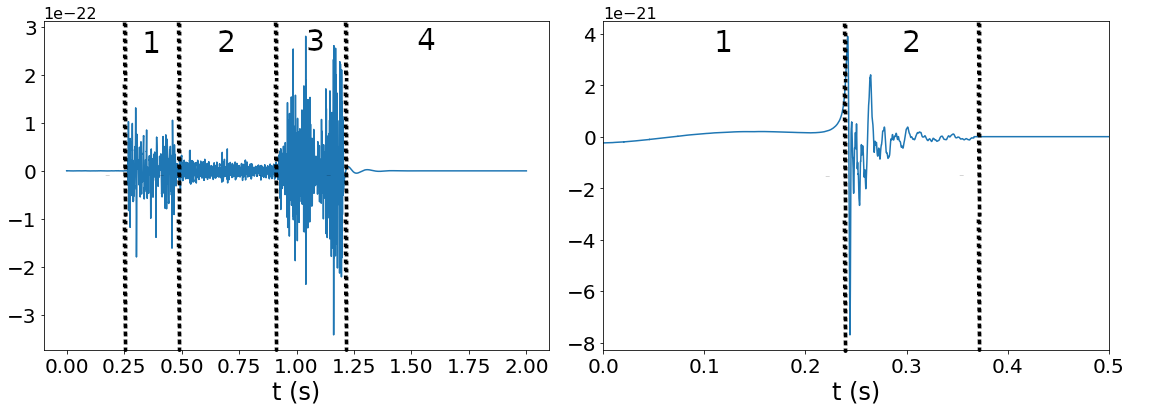} 
\caption{Two examples of the numerically-simulated gravitational waveforms used in this work. The left panel shows a {\tt Mur} signal corresponding to a neutrino luminosity of 6.0 $\times$ 10$^{52}$erg s$^{-1}$ and a 40$M_{\odot}$ progenitor model. The numbers indicate the different phases of the core-collapse process: prompt convection (1), PNS and postshock convection (2), non-linear SASI (3) and explosion onset (4). The right panel shows {\tt Dim} signal {\tt s40A3O09\_LS}, corresponding to a 40$M_{\odot}$ progenitor with strong differential rotation. The waveform displays the pre-bounce phase (1) and the post-bounce PNS ringdown signal (2).}
\label{examplesignals}
\end{figure*}

The idea of the stalled supernova shock being revived by the transfer of energy by neutrinos emitted from the core was first proposed by~\citet{colgate1966hydrodynamic} and~\citet{arnett1966gravitational}, and it was revisited into its modern form by~\citet{bethe1985revival}. During collapse and the subsequent evolution of the PNS into a neutron star, about $\sim 10^{53}$ erg of energy is released~\citep{bethe1990supernova}, mainly in the form of neutrinos. Most of the neutrino emission takes place after bounce, when neutrino losses dominate in the `cooling layer' surrounding the neutrinosphere. Neutrinos can deposit a considerable amount of energy in the `gain layer' between the cooling layer and the shock~\citep{janka2012core}. If a small fraction of this energy is reabsorbed, heating the shock, a successful supernova explosion could ensue. 
Spherically-symmetric simulations implementing neutrino heating have shown that the shock can only be revived in the case of $8-10$~$M_{\odot}$ stars with O-Ne-Mg cores, the lowest-mass massive stars~\citep{kitaura2006explosions}. For more massive progenitors, multidimensional effects, i.e.~convection and SASI, are essential for the success of the neutrino mechanism~\citep{janka2012explosion}. Those nonradial effects increase the efficiency of neutrino heating, and they dominate the GW emission.

In this work we employ the neutrino-driven GW catalog of~\citet{murphy2009model}, which we refer to as the `{\tt Mur}' catalog. It comprises 16 waveforms obtained through Newtonian-gravity 2D simulations accounting for a stalled shock at a radius of $\sim 200$~km, a gain layer beginning at $\sim 100$~km, prompt and postshock convection as well as PNS convection from $\sim20$ to $\sim40$~km and SASI. The effects of neutrino transport are approximated and the simulations account for neutrino cooling and heating as well as the effects of electron capture and deleptonization. The waveforms of~\citet{murphy2009model} correspond to progenitor models with 12, 15, 20 and 40~$M_{\odot}$ zero-age-main-sequence (ZAMS) masses.

The left panel of Figure \ref{examplesignals} shows the waveform of a typical GW signal of the {\tt Mur} catalog. Before bounce, a spherical collapse results in zero GW strain. After the initial bounce, which is marked in the plot as section 1, the stalling shock leaves behind a negative entropy gradient, leading to a convective instability in the gain layer where `prompt convection' takes place. This results in the initial GW burst with peak strains of $\sim 10^{-22}$ (for a source at 10 kpc). Convection takes place both within the PNS and in the postshock region above the neutrinosphere, resulting in GW signals with typical strains of $\sim 10^{-23}$. These are visible once prompt convection settles down as the entropy gradient smooths out, as seen in section 2 of the figure. In the {\tt Mur} catalog this takes place $\sim 50$ ms after bounce~\citep{ott2009gravitational,murphy2009model}. After $\sim 550$ ms SASI reaches a non-linear phase which dominates the signal and causes the amplitude to increase as displayed in section 3 of Fig.~\ref{examplesignals}. During this phase dense plumes from the stalling shock stream down and strike the surface of the PNS, resulting in spikes in the GW signal of several $\sim 10^{-22}$. After this, the explosion begins in section 4. The high-frequency signal is replaced by a low-frequency $\sim 10$~Hz waveform. While this signal would give us information on the shape of the explosion, its detection is unlikely by current GW detectors as those are sensitive to frequencies $\gtrsim20$~Hz. Therefore, the explosion signal would be an abrupt decrease of strain amplitude~\citep{murphy2009model}. Most of the GW signal from the neutrino-driven mechanism is emitted in the $\sim100-1000$~Hz frequency range and it has a duration of $\sim 0.3-1$~s.

\begin{figure*}
\centering
\includegraphics[scale=0.4]{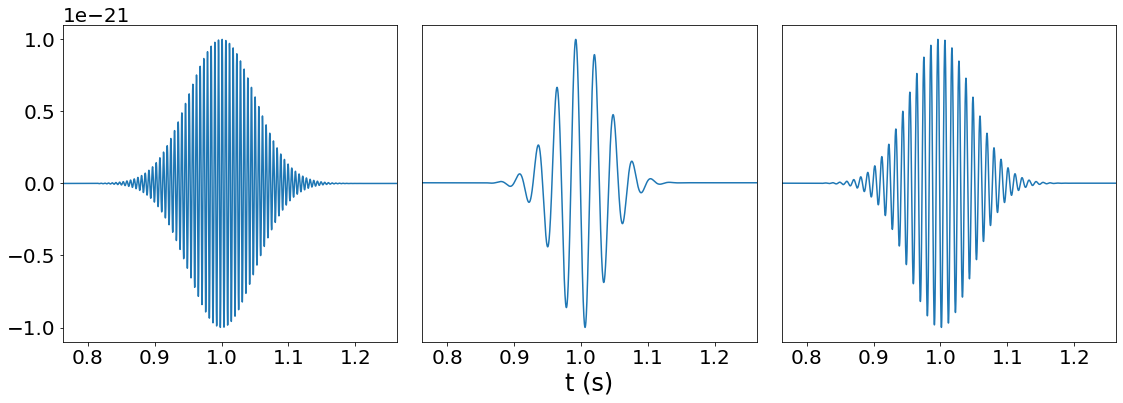} 
\caption{Three randomly chosen sine-Gaussian signals from our generated catalog of 200 signals. They have been rescaled by an order of $10^{-21}$.}
\label{sinegaussians}
\end{figure*}

\subsection{Magneto-rotational mechanism}

MHD phenomena in the context of CCSN were first explored in the 1970s~\citep{bisnovatyi1976magnetohydrodynamic,meier1976magnetohydrodynamic}. Simulations showed that pre-collapse iron cores rotating with periods under $\sim10$~s lead to neutron stars of $\sim1$~ms periods, indicating that conservation of angular momentum leads to a spin-up of the collapsing core by a factor of about $10^3-10^4$. The inner parts of the core collapse further than the supersonically falling outer parts, resulting in a differential rotation. Simulations suggest these rapidly rotating PNS could, $\sim200$~ms after bounce, have a rotational energy of around $10^{52}$~erg, ten times greater than the energy of a typical CCSN~\citep{ott2006spin}.

The magneto-rotational mechanism was first proposed by~\citet{bisnovatyi1971explosion} who suggested that a mechanism involving magnetic fields would be able to efficiently transfer angular momentum so that the core's rotational energy could be used to revive an explosion. Simulations later found that this mechanism was indeed efficient enough to extract enough energy to trigger a CCSN explosion~\citep{moiseenko2005magnetorotational}. The non-linear magneto-rotational instability (MRI) was suggested as the dominant mechanism to amplify magnetic fields to the necessary magnitudes to drive an explosion~\citep{akiyama2003magnetorotational}. Indeed, in simulations where MRI was not present and magnetic field amplification ocurred solely due to collapse and wrapping of the field lines it was found that only progenitors with initially strong magnetic fields ($\sim10^{12}-10^{13}$~G) could launch an explosion~\citep{sawai2005core}. Simulations accounting for MRI, however, suggest that MRI after bounce is the main driver of field amplification, working for even initially weak fields~\citep{moiseenko2006magnetorotational,obergaulinger2006axisymmetric,kotake2004magnetorotational,Kuroda:2020,Jardine:2021,Martin:2020}. These simulations all found collimated, jet-like explosions.

Our rapidly-rotating waveform signals were extracted from the waveform catalog of~\citet{dimmelmeier2008gravitational}, which we will refer to as the `{\tt Dim}' catalog. It consists of 128 signals obtained through 2D simulations accounting for general relativity, two different equations of state (EoS), approximated deleptonization and neutrino pressure and different initial rotation values and degrees (uniform and differential). The catalog comprises progenitor models with 11.2, 15, 20 and 40~$M_{\odot}$ ZAMS masses.

The right panel of Figure \ref{examplesignals} shows a representative waveform of the {\tt Dim} catalog. Simulations have shown GW from rotating CCSN have a generic morphology, known as Type I signals~\citep{dimmelmeier2008gravitational}: a steady rise pre-bounce, marked as section 1 in the plot, followed by a strong negative peak at bounce, corresponding to the rapidly changing quadrupole moment of the inner core which is deformed by centrifugal forces. Milliseconds after bounce, convection sets in, resulting in a ringdown signal shown in section 2 \citep{dimmelmeier2008gravitational,ott20073d,dimmelmeier2007generic}. The frequency of most of the GW signals emitted are in the range $\sim 500-800$~Hz and the signals have a duration of several $\sim10$~ms after bounce.

\subsection{Sine-Gaussian signals}
  
In addition to the two types of signals we have just discussed, we will also employ a third `mock' catalog. This catalog comprises simulated signals modeled as sine-Gaussian (SG) waveforms, which serve the purpose of estimating detection sensitivity to generic, short-duration GW signals. The main motivation in introducing this catalog is to expand the number of signal types we can apply our classification approach to, which allows us to further assess the algorithm. We also note that \textit{ad hoc} signals such as this type of waveforms have been used by the LVK in optically-targeted CCSN searches to estimate the energy constraints of GW emitted by such source~\citep{abbott2020optically}. We build this catalog, dubbed `{\tt SG}', by creating an array of 200 sine-Gaussian signals, varying in duration from 0.05 to 0.1~s and in frequency from 20 to 200~Hz, as well as having a varying phase. Figure \ref{sinegaussians} shows three illustrative examples of waveforms from this catalog.

\section{Signal reconstruction and classification}

We turn now to explain the two-step process we have developed to reconstruct and classify GW signals. The interested reader is addressed to~\cite{torres2016denoising} for further details.

\subsection{Fundamentals}

In our approach we define $\textbf{D}\in \mathbb{R}^{l\times p}$ as the so-called dictionary, a matrix composed of $p$ atoms of length $l$, and $\boldsymbol\alpha \in \mathbb{R}^p$, which is a vector of coefficients that represents the reconstructed signals. This vector should be as sparse as possible, that is, it should contain a small amount of non-zero coefficients. A variational problem balances the reconstruction of the original signals ${\textbf{u}}$ while ensuring that vector $\boldsymbol\alpha$ contains as many zeros as possible. This variational problem is known as basis pursuit~\citep{chen2001atomic} or LASSO~\citep{tibshirani1996regression}, and reads
\begin{equation}
\label{eqn:lasso1}
\boldsymbol\alpha= \argmin{(\norm{\textbf{D} \boldsymbol\alpha - \textbf{f}}^2_2+\lambda\norm{\boldsymbol\alpha}_1)}\,,
\end{equation} 
where 
$\textbf{f}$ is a vector of noisy signals resulting from injecting signals $\textbf{u}$ into Gaussian noise $\textbf{n}$, and 
$\norm{\cdot}_1$ and $\norm{\cdot}_2$ are the L$_1$-norm and the L$_2$-norm, respectively. The L$_1$-norm term introduces a Lagrange multiplier $\lambda$, or regularization parameter, which helps ensure a sparse $\boldsymbol\alpha$ and avoids overfitting. Large values of $\lambda$ would return a coefficient vector with zero L$_1$-norm. If $\lambda=0$, the variational problem would only contain the first term and would simply correspond to a least-square regression problem. 

In this work we use two types of dictionaries. First, we separate each waveform catalog into two sets, roughly split in a percentage 80/20. While the smaller sets consist of the signals we will attempt to denoise and reconstruct, the bigger sets will be directly used as dictionaries without any prior training (see below), so that each atom is one of the original catalog waveforms. We refer to these dictionaries as $\textbf{D}^{\rm cat}$ and they will be used to classify signals by their type. The other class of dictionary, $\textbf{D}^{\rm train}$, is trained using patches extracted from the same 80\% set. These will be the dictionaries used for signal denoising. The dictionary training method is explained in Section 4.

Each signal will undergo two reconstruction processes. First, the denoising is done over learned dictionaries, an effective method to solve signal-processing problems (see~\citet{mairal2009online} and references therein). This begins by extracting patches from an initial noisy signal \textbf{f} and solving the LASSO problem for $\boldsymbol\alpha$ over a previously trained dictionary, $\textbf{D}^{\rm train}$. Then, a denoised signal \textbf{y} is built out of the patches obtained from the product of $\textbf{D}^{\rm train}$ and $\boldsymbol\alpha$. The second reconstruction process, pertaining to the signal classification process, is done over the waveform dictionaries $\textbf{D}^{\rm cat}$. Each denoised signal $\textbf{y}$ is reconstructed whole, rather than out of patches, as we aim to find out which waveforms best reconstruct a denoised signal.

Carrying out these two processes, from calculating the best denoising parameters to determining which catalog best reconstructs the signals, necessitates some way of quantifying which reconstructions actually correspond to the best results. We will define the best results as those that minimize the Wasserstein distance (WD), a metric function which measures the minimum amount of `work' required to transform $\textbf{u}$ into the reconstructed signal~\citep{ramdas2017wasserstein}. To compute WD in specific ranges of the time series waveforms must be aligned. All 128 {\tt Dim} signals are aligned at core bounce, that is, the large negative peak immediately after the pre-bounce rise. As can be seen in the right panel of Figure \ref{examplesignals}, this happens at 0.24~s, though this will later be changed to 0.74~s as we will pad the signals to be a second longer. Most of the {\tt Mur} signals are contained in the 0.15$-$1.29 s range. Correspondingly, {\tt SG} signals are centered at 1 s and vary between 0.05 and 0.1 s in duration, as shown in Figure \ref{sinegaussians}. 

A flowchart of the entire two-step process we use is shown in Fig.~\ref{diagrama}. First, the three catalogs are split into bigger sets, $\textbf{D}^{\rm cat}$, and smaller sets of $\textbf{u}$ signals. From the former we extract training patches and find the optimal parameters to train $\textbf{D}^{\rm train}$ dictionaries through the LASSO variational problem. 
Those optimal parameters are determined by (i) injecting the $\textbf{u}$ signals into Gaussian noise to obtain $\textbf{f}$ signals, (ii) denoising the latter with all trained dictionaries, (iii) 
computing the WD between each original signal $\textbf{u}$ and the corresponding denoised signal $\textbf{y}$, and (iv) choosing the parameters that minimize said WD. The dictionary-training part of the process assumes that we know to which class the original signals belong to, so that e.g.~only $\textbf{D}^{\rm train}_{\rm Dim}$ is used to denoise $\textbf{f}_{\rm Dim}$ signals and so on. Further details of each of the two steps are outlined next.

\subsection{Reconstruction step}

As mentioned before, each of the {\tt Dim}, {\tt Mur}, and {\tt SG} catalogs is separated into two 80/20\% sets. We denote the larger sets $\textbf{D}_{\rm Dim}^{\rm cat}$, $\textbf{D}_{\rm Mur}^{\rm cat}$ and $\textbf{D}_{\rm SG}^{\rm cat}$, as they can themselves be used as dictionaries. The first thing we have to do with them is to extract $m$ training patches from all the signals available in the sets. From these patches we train dictionaries $\textbf{D}_{\rm Dim}^{\rm train}$, $\textbf{D}_{\rm Mur}^{\rm train}$ and $\textbf{D}_{\rm SG}^{\rm train}$, a process mediated by the regularization parameter $\lambda^{\rm train}$ (see Section 4).

\begin{figure}
\centering
\includegraphics[scale=0.28]{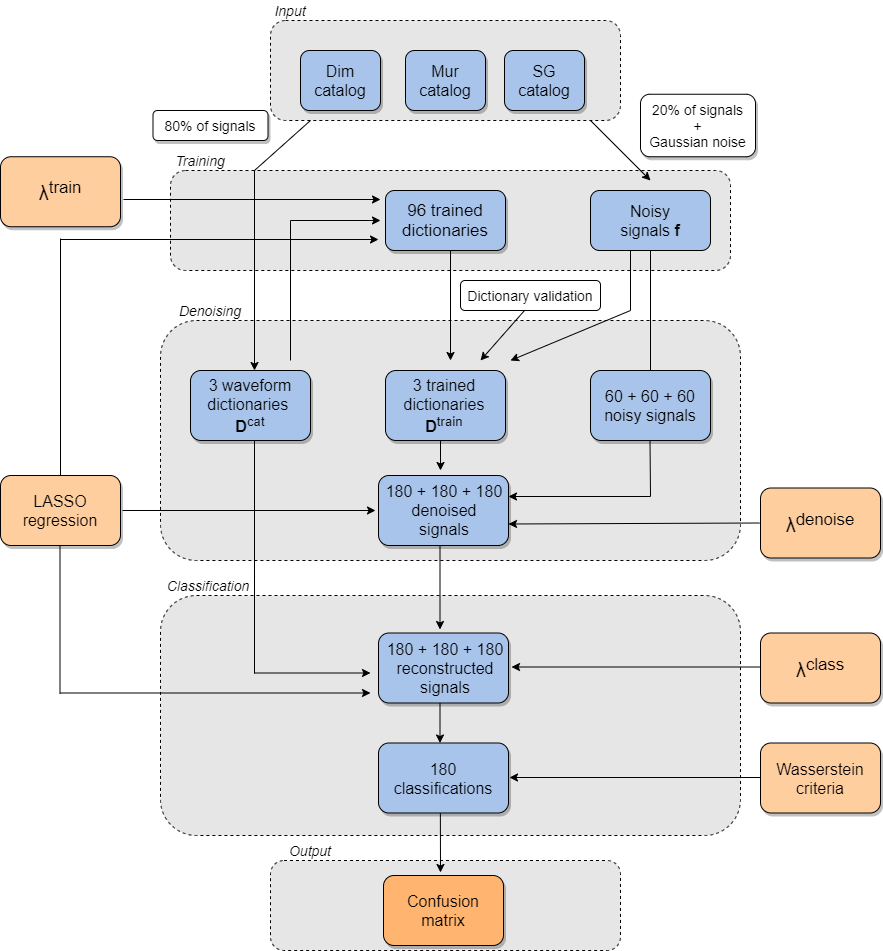} 
\caption{Flow diagram of the two-step process to reconstruct and classify GW signals using learned dictionaries.}
\label{diagrama}
\end{figure}

The signals $\textbf{u}$ of the smaller sets are injected into Gaussian noise $\textbf{n}$ to produce signals $\textbf{f}=\textbf{u}+\textbf{n}$. The first operation we perform on these noisy signals is to denoise (reconstruct) them using the previously trained dictionaries $\textbf{D}^{\rm train}$. Therefore, 
denoising $\textbf{f}$ with $\textbf{D}^{\rm train}_{\rm Dim}$ will yield a $\textbf{y}_{\rm Dim}$ signal and so on. This reconstruction step is done following the LASSO problem 
where $m$ patches will have been extracted from the noisy signals. We denote the regularization parameter of this problem $\lambda^{\rm denoise}$. One of the preliminary steps of our procedure is therefore to find an optimal value of $\lambda^{\rm denoise}$ for each signal type as well as the optimal parameters of the learned dictionaries. This corresponds to the `Denoising' section of the flowchart in Fig.~\ref{diagrama}.

%

\subsection{Classification step}

The classification step is carried out using the three denoised signals $\textbf{y}_{\rm Dim}$, $\textbf{y}_{\rm Mur}$, and $\textbf{y}_{\rm SG}$. This allows us to identify whether $\textbf{u}$ is a signal resulting from a magneto-rotational-driven CCSN ({\tt Dim}), a neutrino-driven CCSN ({\tt Mur}), or an \textit{ad-hoc} sine-Gaussian signal ({\tt SG}). For this purpose we work with 
the entire denoised signal $\textbf{y}$ instead of working with patches as done during the denoising step. We denote the regularization parameter of this process $\lambda^{\rm class}$ and the dictionaries we employ will now be the 80\% sets $\textbf{D}^{\rm cat}$. As $\boldsymbol\alpha$ is now a vector of size $p$, where $p$ is the number of waveforms present in $\textbf{D}^{\rm cat}$, each non-zero coefficient $\boldsymbol\alpha_i$ indicates the degree to which $\textbf{y}$ is reconstructed from the $i$-th waveform in the catalog. We will show below some results to illustrate this.

Signals $\textbf{y}_{\rm Dim}$, $\textbf{y}_{\rm Mur}$ and $\textbf{y}_{\rm SG}$, obtained using each of the three trained dictionaries, are our first set of reconstructions. Our second set is obtained by carrying out the regression step. Thus, reconstructing $\textbf{y}_{\rm Dim}$ over $\textbf{D}^{\rm cat}_{\rm Dim}$ will yield $\textbf{y}_{\rm Dim-Dim}$ (and similarly for $\textbf{y}_{\rm Mur-Mur}$ and $\textbf{y}_{\rm SG-SG}$). While a more in-depth analysis can be carried out through cross reconstructions, such as $\textbf{y}_{\rm Mur-Dim}$, here we will only work with signals reconstructed with matching dictionary types.
We will then compute the WD between $\textbf{y}_{\rm Dim} - \textbf{y}_{\rm Dim-Dim}$, $\textbf{y}_{\rm Mur} - \textbf{y}_{\rm Mur-Mur}$ and $\textbf{y}_{\rm SG} - \textbf{y}_{\rm SG-SG}$, as in a realistic situation we will not have access to the original signal $\textbf{u}$. Our classification criterion considers that the smallest WD value will determine whether $\textbf{u}$ is either a {\tt Dim}, {\tt Mur}, or {\tt SG} signal. When presenting our results we will go over the ranges we choose to compute the WD since, as we shall see, those may affect the results, particularly in the case of {\tt Mur} waveforms.

By doing this classification process with a number of different signals and contrasting the results with the true nature of said signals, we are able to build a confusion matrix that serves to assess the effectiveness of our algorithm. 

\section{Dictionary training}

\subsection{Separating signals and training the dictionaries}

Waveforms from the catalogs are separated into signals used to train the dictionaries and signals to validate the approach through reconstruction and classification. For the {\tt Dim} catalog we choose 102 signals to train and 26 signals for validation, for the {\tt Mur} catalog 11 and 5 signals, respectively, to account for their small number, and for the {\tt SG} catalog 160 and 40 signals, respectively. While the {\tt SG} simulated signals are generated with an amplitude between $-1$ and 1, the 102 and 11 sets of {\tt Dim} and {\tt Mur} signals have to be first rescaled between $-$1 and 1. This is necessary to avoid the difficulties of training dictionaries out of patches with $10^{-22}-10^{-21}$ amplitudes.

Once this is done we extract $m$ patches of length $n$ from each of the three larger sets. The function designed to do this lets us specify how much overlap is allowed, which in this case we set to maximum as to allow us to extract a sufficiently high number of patches. It is also designed to avoid extracting patches of zero module, which would interfere with the dictionary-training process. Furthermore, we normalize these patches to zero mean and unit standard deviation. For all cases we work with $m=3000$ training patches and four different values of $n$. Specifically we work with patches of size $n=64, 128, 256$ and 512 for the {\tt Dim} and {\tt Mur} dictionaries while for {\tt SG} signals we extract patches with $n=128, 256, 512$ and 1024 instead, as patch sizes smaller than that resulted problematic when it came to training the dictionaries for this particular type of signals.

Before training the dictionaries from these patches, the smaller sets of signals are injected into Gaussian noise. Both the {\tt Dim} and {\tt Mur} catalogs have a sampling frequency of 16384 Hz but they respectively have a duration of 1 s and 2 s. As our final objective is to reconstruct a signal of unknown origin, we pad the {\tt Dim} signals with zeros so that every signal we work with consists of 32768 samples. Once all signals have the same length we generate Gaussian noise by specifying the duration and sampling frequency of the signals as well as a cutoff frequency of 20~Hz, according to the $\sim$10$-$20~Hz sensitivity lower limit of the Advanced LIGO and Advanced Virgo detectors \citep{sengupta2016sensitivity,abbott2020prospects}. We also employ the sensitivity curve of Advanced LIGO in the form of the power spectral density (PSD) of noise sources~\citep{aligo}.  

\begin{figure}
\centering
\includegraphics[scale=0.21]{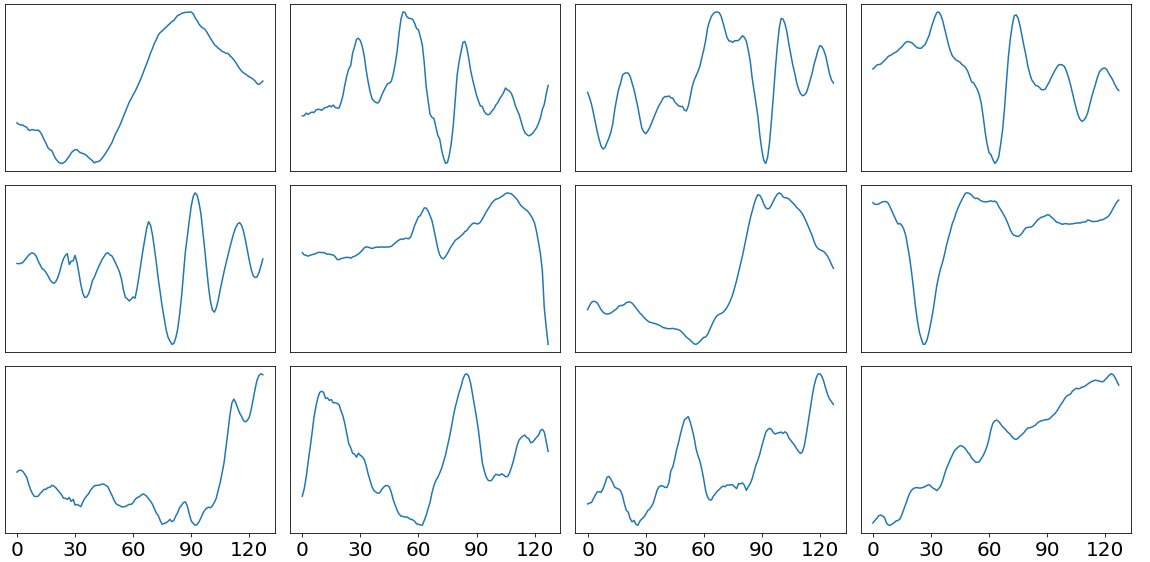}
\\
\vspace{0.3cm}
\includegraphics[scale=0.21]{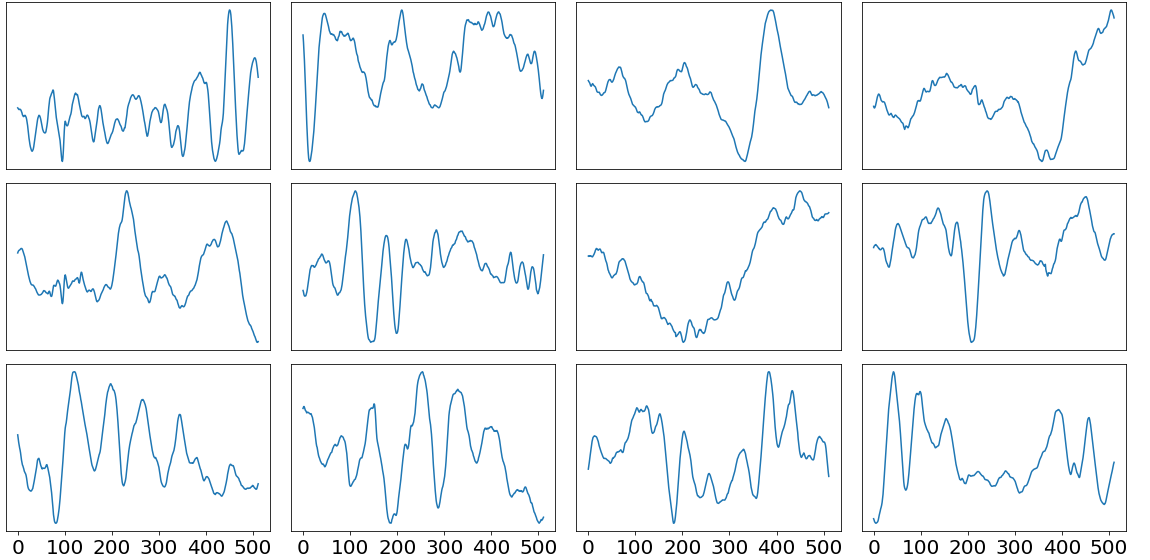} 
\\
\vspace{0.3cm}
\includegraphics[scale=0.21]{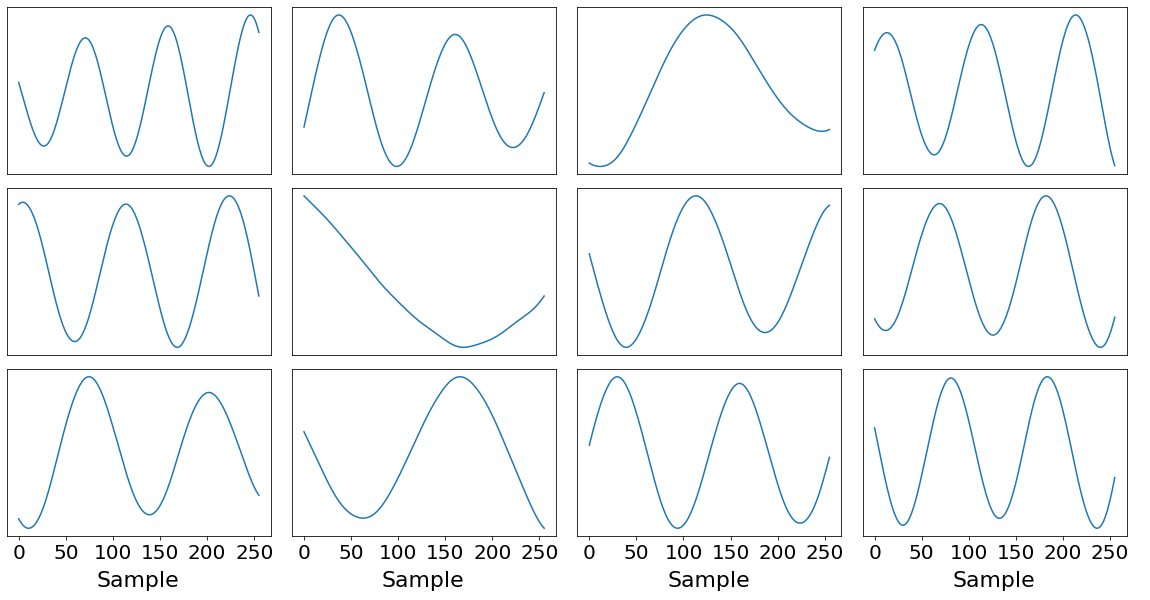} 
\caption{Examples of several atoms from three trained dictionaries. The top panel corresponds to a {\tt Dim} dictionary trained with $n=128$ and $\lambda^{\rm train}=1.2$. The plots in the central panel show atoms from a {\tt Mur} dictionary trained with $n=512$ and $\lambda^{\rm train}=1.7$. The bottom panel corresponds to a {\tt SG} dictionary trained with $n=256$ and $\lambda^{\rm train}=0.7$. All of the atoms in these examples have been extracted from dictionaries with $p=100$.}
\label{dictexamples}
\end{figure}

To inject each signal into Gaussian noise we need to scale its amplitude to a specific signal-to-noise ratio (SNR). The SNR is defined as
\begin{equation}
\textrm{SNR} = \sqrt{4\Delta t^2 \Delta f \sum_{k=1}^{N_f}\frac{|\tilde{h}(f_k)| ^2}{S(f_k)}},
\end{equation}
where $\tilde{h}$ is the Fourier transform of the signal strain $h$, $f_k$ are the elements of the frequency vector, $S(f_k)$ is the PSD of the noise, $N_f$ is the number of frequencies, and $\Delta t$ and $\Delta f$ are the time and frequency steps, respectively. In this work we will study the dependence of our signal classification procedure with the SNR. We will use the values SNR=20, 15, 10 and 5, the latter being a fairly low SNR scenario.

We turn now to discuss the training of the dictionaries using the large sets of signals from the {\tt Dim}, {\tt Mur}, and {\tt SG} catalogs. A key parameter of the LASSO problem is the regularization parameter. We start analysing how to set the lower and upper limits of this parameter, $\lambda^{\rm train}$, to train dictionaries, referring to the next section the discussion of which of the parameters best denoise signals. If $\lambda^{\rm train}$ is too large then the minimization problem of training a dictionary will prioritize solutions with zero L$_1$-norm, resulting in noisy dictionary elements which would difficult the reconstruction of signals~\citep{torres2016denoising}. On the other hand, if $\lambda^{\rm train}$ is too small then the dictionary training problem will turn into a least-square regression problem, resulting in an overfitting of the dictionary elements which would come with a loss of generality. 

\begin{figure}
\centering
\includegraphics[scale=0.21]{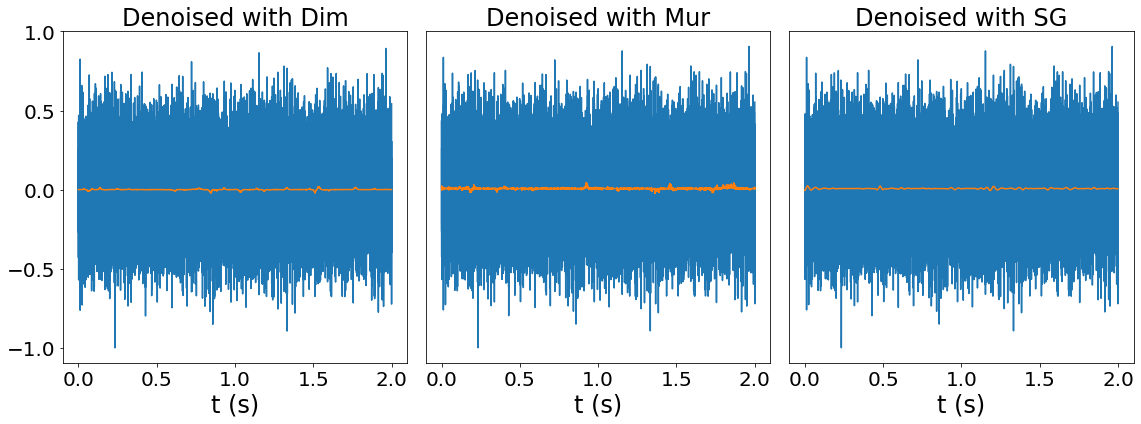} 
\caption{The orange curves show the results of denoising pure Gaussian noise (blue; no signal injected) using the $\textbf{D}^{\rm train}$ dictionaries, {\tt Dim} (left), {\tt Mur} (centre), and {\tt SG} (right).}
\label{purenoise}
\end{figure}

For every signal type we generate dictionaries with four different values of $n$, as previously mentioned. In addition, for each $n$ we generate two dictionaries, each with a different number of atoms, $p=100$ and $p=200$, which gives eight different combinations of $n$ and $p$ for each signal type. Furthermore, once the lower and upper limits of $\lambda^{\rm train}$ are selected for each signal type and $n$, we train each of these dictionaries with 4 different values of $\lambda^{\rm train}$. Therefore, in total we build 32 dictionaries for each signal type. As an illustration, several atoms from three of these dictionaries are shown in Figure~\ref{dictexamples} for the three types of waveform signals used in this study. Technical details on the computation and optimization of the parameters of the dictionaries are presented in Appendix A. In particular, the optimal values of $\lambda^{\rm train}$ we obtain, i.e.~those that correspond to the smallest WD, for signals with SNR=20, are reported in Table~\ref{lambdatrainopt}.

If the denoising is carried out over pure Gaussian noise, all three trained dictionaries create spurious signals where there should be none. This is illustrated in Figure \ref{purenoise} where the spurious reconstruction is compared with the original noise. Even though a fake reconstruction appears, its amplitude is small. These artificial signals could be reduced by increasing the value of $\lambda^{\rm denoise}$. However, doing so in general might lead to missing actual low SNR GW injections. Therefore, our final denoising and classification results, discussed in Section 5, are obtained with the optimal values of $\lambda^{\rm train}$ and $\lambda^{\rm denoise}$ reported in Table~\ref{dictionaryparameters} of Appendix A.  

\subsection{Waveform dictionaries}

\begin{figure}
\centering
\includegraphics[scale=0.21]{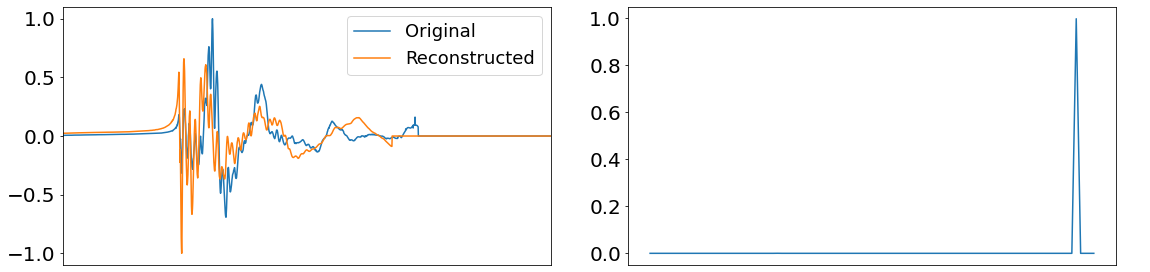}
\includegraphics[scale=0.21]{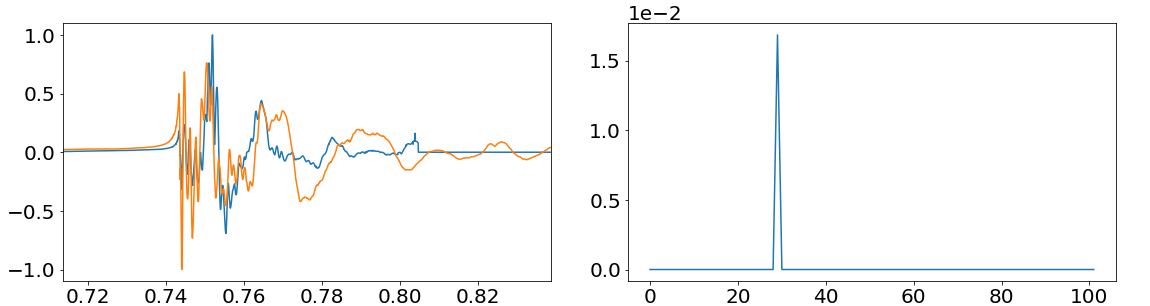} 
\\
\vspace{0.5cm}
\includegraphics[scale=0.21]{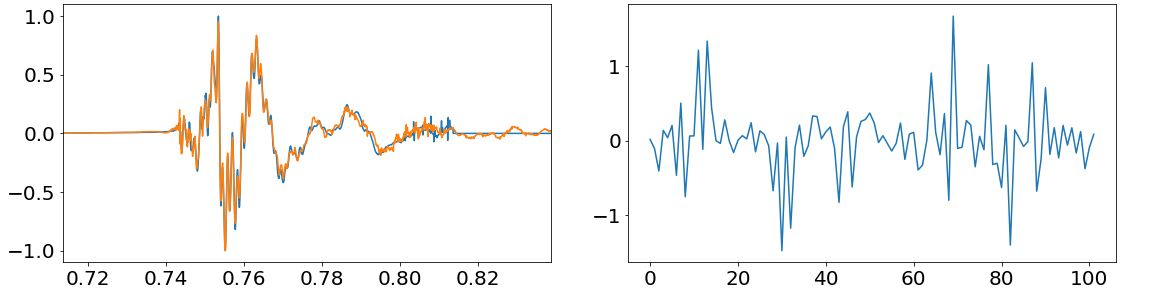}
\includegraphics[scale=0.21]{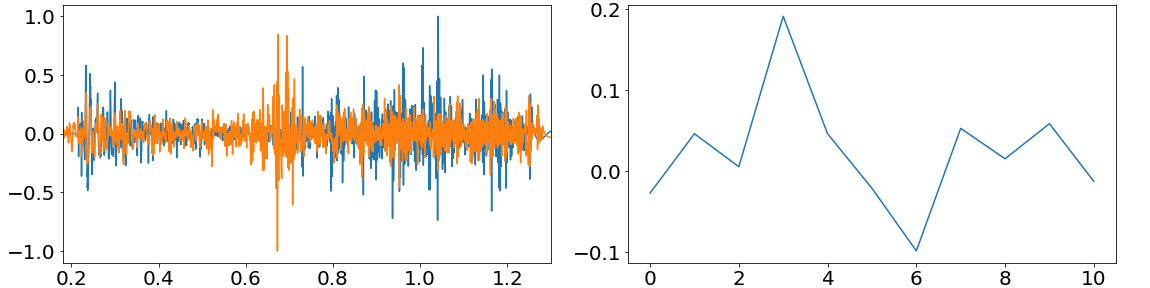} 
\includegraphics[scale=0.21]{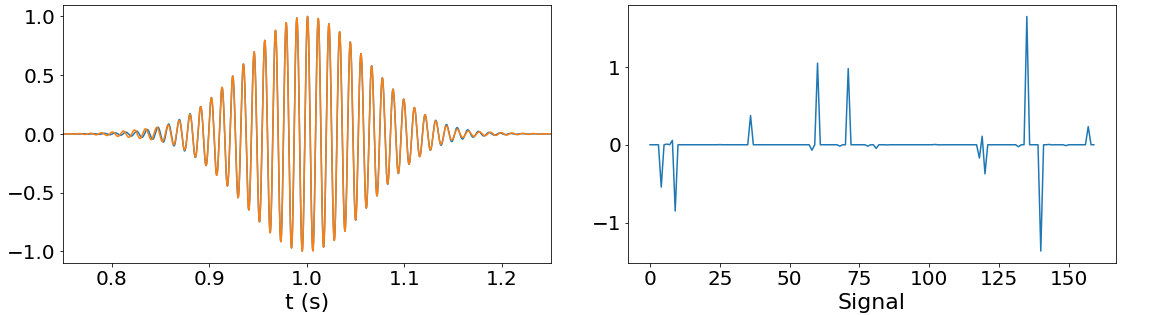} 
\caption{Left panels: reconstructions of several representative signals using $\textbf{D}^{\rm cat}$ dictionaries. Right panels: corresponding values of coefficient vector $\boldsymbol\alpha$. The first row shows the reconstruction of the noiseless signal no.~97 of the {\tt Dim} catalog using the $\textbf{D}^{\rm cat}_{\rm Dim}$ dictionary at its optimal $\lambda^{\rm class}$ , while the second row showcases the reconstruction of that same signal while embedded in noise with SNR=20 at a $\lambda^{\rm class}$ which forced one non-zero coefficient. The next three rows show the optimized reconstructions of noiseless {\tt Dim}, {\tt Mur}, and {\tt SG} signals (from top to bottom) which are not contained in their respective $\textbf{D}^{\rm cat}$ dictionaries.}
\label{alphas}
\end{figure}

We now discuss the dictionaries that will be used for signal classification, $\textbf{D}^{\rm cat}$. The classification is done by solving the LASSO variational problem over previously denoised signals. Figure \ref{alphas} shows the result of reconstructing signals using $\textbf{D}^{\rm cat}$ instead of $\textbf{D}^{\rm train}$. The left panels compare the original and reconstructed signals and the right panels indicate the non-zero values of the coefficient vector ${\boldsymbol\alpha}$.
The upper row shows a {\tt Dim} signal which is contained in the dictionary, namely signal 97 of $\textbf{D}^{\rm cat}_{\rm Dim}$. Therefore, ideally the reconstruction should result in only one non-zero $\boldsymbol\alpha$ coefficient which exactly matches the signal. This is what happens when reconstructing a signal without noise, as can be seen in the right panel of the first row. The situation is, however, quite different when reconstructing that same signal embedded in noise, as shown in the second row for the particular case SNR=20 (note that we do not plot the noise in the left panel of the figure). In this case, if we force the algorithm to accept only a reconstruction with one non-zero coefficient, it does not always return the correct one, as seen in the right panel. If the previous condition is not enforced, the reconstruction results in many non-zero coefficients.

This last outcome is also what happens in a situation more similar to what can be found in practice, namely when the waveform dictionaries are used to reconstruct signals that are not contained in them. In this case we allow the signals to be reconstructed with the number of non-zero coefficients which optimizes the WD. The results are plotted in the lower three rows of Fig.~\ref{alphas} for {\tt Dim}, {\tt Mur}, and {\tt SG} signal (from top to bottom). As a test, the results shown in this figure are performed over noiseless signals. The reconstructions shown yield the following values of our metric estimator: WD$_{\rm Dim} = 8.99 \times 10^{-3}$, WD$_{\rm Mur} = 1.36 \times 10^{-2}$ and WD$_{\rm SG} = 1.37 \times 10^{-3}$. As can be seen both by these values and qualitatively (in the signal comparison in the left panels), the {\tt SG} and {\tt Mur} reconstructions give, respectively, the best and worst results. This is to be expected as catalog $\textbf{D}^{\rm cat}_{\rm SG}$ has the largest number of signals to draw from, at 160, while $\textbf{D}^{\rm cat}_{\rm Mur}$ has the smallest, at 11.

\section{Results}

\subsection{Waveform classification}

For each SNR we build four different confusion matrices. We work with 180 signals for each matrix, 60 from each catalog, extracted from the small sets of (validation) signals not included in the dictionaries. As we only have 26, 5 and 40 {\tt Dim}, {\tt Mur} and {\tt SG} signals available, respectively, many of them will be repeated, although they are chosen randomly and they are injected each time in newly generated noise realizations.

\begin{figure}
\centering
\includegraphics[scale=0.20]{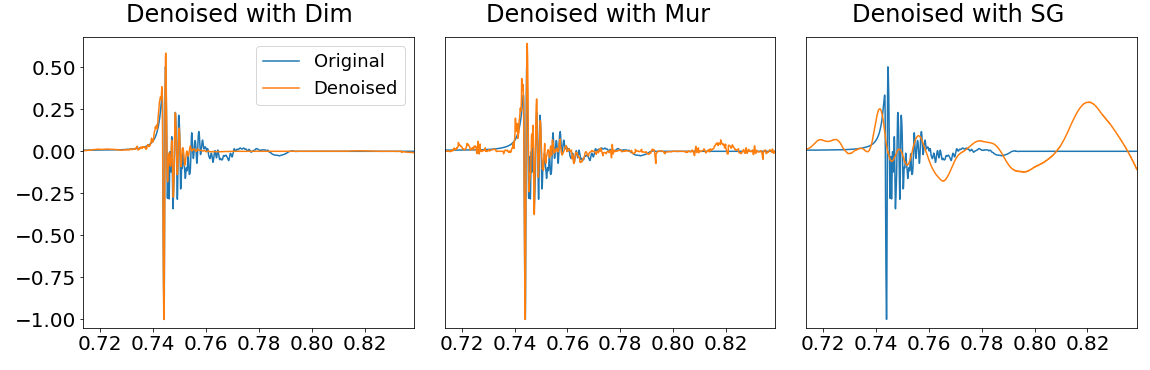}
\includegraphics[scale=0.20]{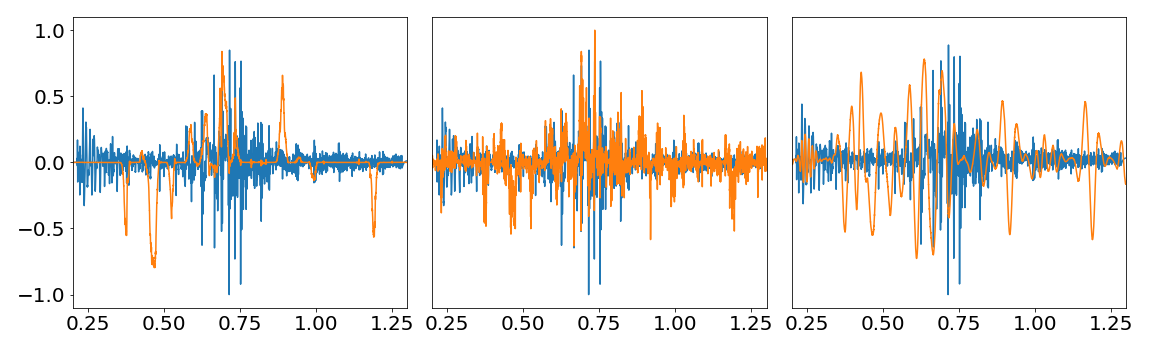}
\includegraphics[scale=0.20]{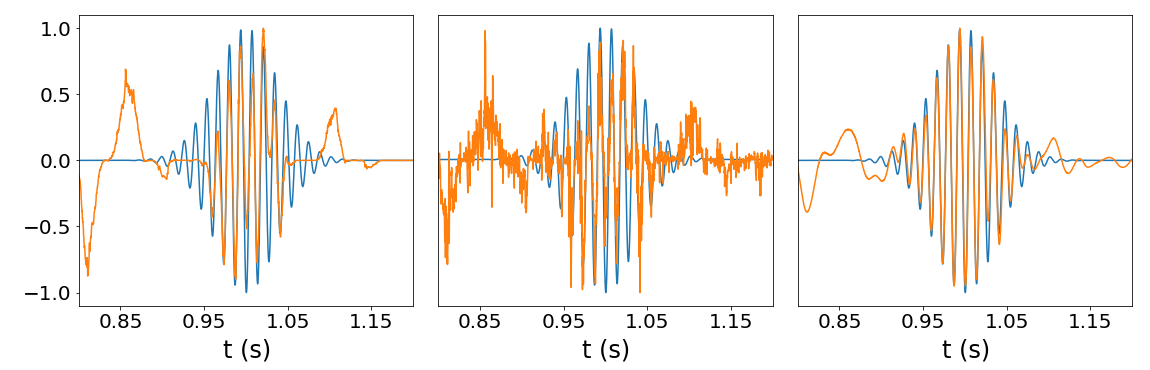} 
\caption{Results of denoising three representative SNR=20 signals, {\tt Dim} (top), {\tt Mur} (centre), and {\tt SG} (bottom), using all three learned dictionaries $\textbf{D}^{\rm train}$.}
\label{ninedenoisings}
\end{figure}

We start by denoising these signals with all three different $\textbf{D}^{\rm train}$ using the parameters of Table~\ref{dictionaryparameters} and then we reconstruct these denoised signals using the waveform dictionary of the same type. Therefore, each matrix is created out of 540 reconstructions: 180 signals reconstructed first with $\textbf{D}^{\rm train}_{\rm Dim}$ and then with $\textbf{D}^{\rm cat}_{\rm Dim}$, and the same amount for the {\tt Mur} and {\tt SG} catalogs. The second LASSO regression is performed with the value of $\lambda^{\rm class}$ that optimizes the WD between the reconstructed and the denoised signals for each individual case. As explained in Section 3.3, we limit our study to reconstructions created with dictionaries of the same type. For illustration purposes Figure~\ref{ninedenoisings} shows the outcome of denoising one example of each type of signal with either $\textbf{D}^{\rm train}$ for SNR=20. As expected, each signal type is denoised best with its respective dictionary.
 
Therefore, each original signal $\textbf{f}$ will result in three final reconstructions, $\textbf{y}_{\rm Dim-Dim}$, $\textbf{y}_{\rm Mur-Mur}$ and $\textbf{y}_{\rm SG-SG}$. Associated with each one of these there is a WD obtained by the comparison with the denoised signals $\textbf{y}_{\rm Dim}$, $\textbf{y}_{\rm Mur}$ and $\textbf{y}_{\rm SG}$. We choose the best reconstruction as the one that minimizes the WD and by comparing that reconstruction with the actual signal $\textbf{u}$, we are able to build the components of a confusion matrix.

\begin{figure}
\centering
\includegraphics[scale=0.20]{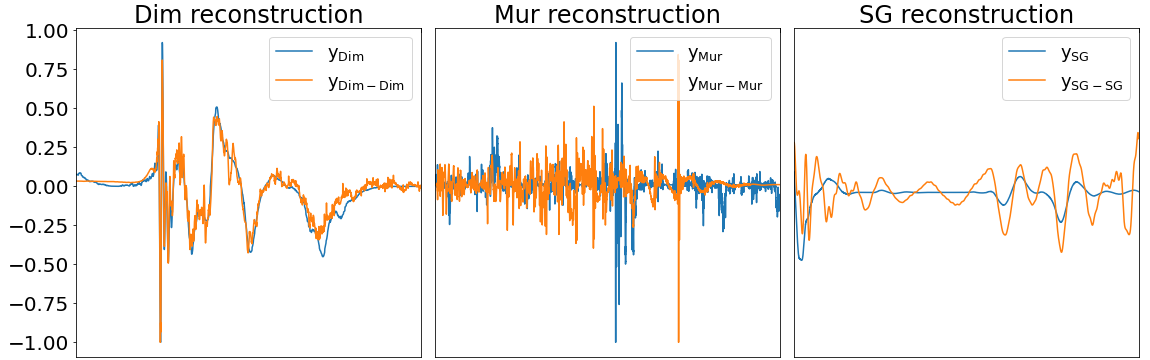} 
\includegraphics[scale=0.20]{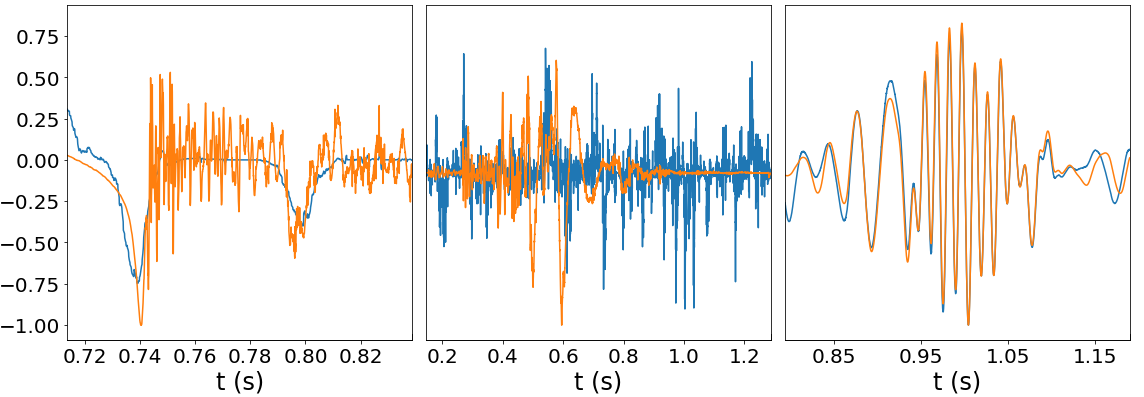} 
\caption{Reconstructions of {\tt Dim} signal 99 at SNR=20 (top panels) and {\tt SG} signal 23 at SNR=10 (bottom panels). Computing the three WD in the ranges pictured in the plots results in both signals being misclassified as {\tt Mur}.}
\label{recfallidas}
\end{figure}

It is necessary to explain the criterion we use to choose the best reconstruction based on the WD. While the optimization process to calculate the regularization parameters computes the WD in the ranges where most of the signals are located, that does not produce satisfactory results in the classification process. At SNR=20, 92\% of the {\tt Dim} signals are correctly identified even when qualitatively those 8\% misidentified signals show a successful reconstruction of a {\tt Dim} signal, as depicted in the top panels of Figure~\ref{recfallidas}. Even more of a concern is the fact that at SNR=10 the percentage of correctly identified {\tt SG} signals falls to only 20\%. The bottom panels of Figure~\ref{recfallidas} show a clear reconstruction of a {\tt SG} signal being misidentified as a {\tt Mur} signal (shown in blue).

Focusing on the top panels of Figure \ref{recfallidas}, $\textbf{y}_{\rm Mur}$ (shown in blue) does show a large peak corresponding to the bounce peak of {\tt Dim} signals in the 0.71$-$0.84 s range. This is not a surprise as Figure \ref{ninedenoisings} already shows that $\textbf{D}^{\rm train}_{\rm Mur}$ does an acceptable job of denoising a {\tt Dim} signal. Despite this, calculating the WD of the {\tt Dim} reconstruction in the 0.71$-$0.84~s range (top-left plot of Fig.~\ref{recfallidas}) and of the {\tt Mur} reconstruction in the 0.15$-$1.29~s range (top-centre plot), leads to WD$_{\rm Dim}=0.062$ and WD$_{\rm Mur}=0.022$. Therefore, the signal is misclassified as {\tt Mur} by having a smaller WD. This could be due to the fact that {\tt Mur} signals are stochastic and they are the longest of the three catalogs, which may impact the way the WD quantifies the work needed to turn one signal into another. We explored the possibility of using other similarity indices, such as the mean-squared error. In this case, the classification of {\tt Dim} and {\tt SG} signals was robust but only about 40\% of {\tt Mur} signals were correctly identified at SNR=20.

To mitigate those issues we employ the WD in the following way: First, 
we compute WD$_{\rm Dim}$ and WD$_{\rm Mur}$ in the temporal range of {\tt Dim} signals (i.e.~in the 0.71-0.84 s range in the example of Fig.~\ref{recfallidas}). By doing this we compare two signals of equal length and we give prominence to identifying whether the distinctive bounce peak of {\tt Dim} signals is present. If WD$_{\rm Dim}<{\rm WD}_{\rm Mur}$, we compare WD$_{\rm Dim}$ to WD$_{\rm SG}$ in the temporal range of {\tt SG} signals (namely, in the 0.79-1.19 s range in the example of Fig.~\ref{recfallidas}). If instead WD$_{\rm Dim}> {\rm WD}_{\rm Mur}$, we recalculate WD$_{\rm Mur}$ in the {\tt SG} signals range to compare it to WD$_{\rm SG}$. In either case we choose the smallest value of the WD. This method correctly classifies the two signals of Fig.~\ref{recfallidas}. We note that while the procedure has been illustrated with only a few signals we apply it to the full 180 signals of each confusion matrix.

\begin{figure}
\centering
\includegraphics[scale=0.21]{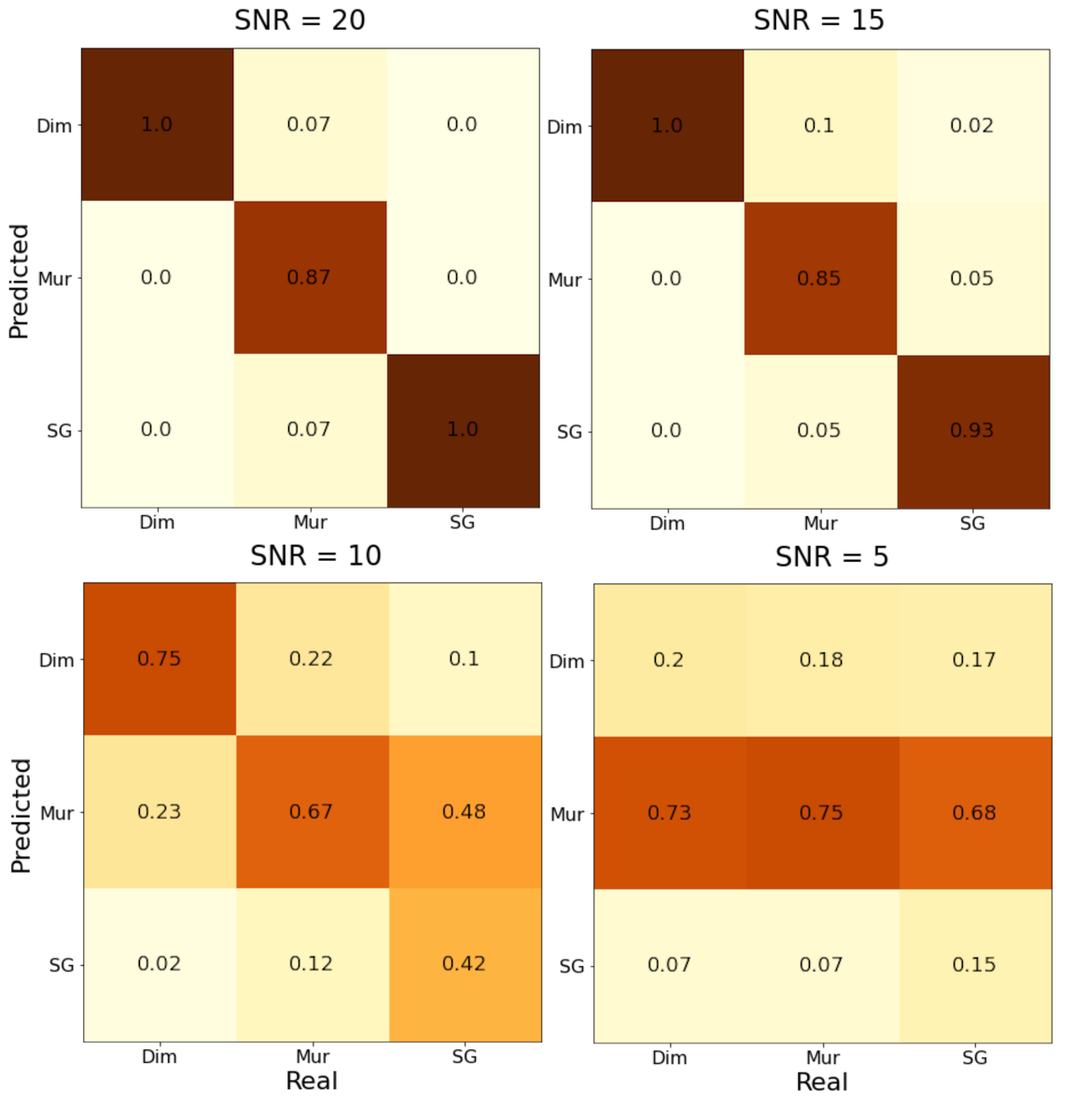} 
\caption{Normalized confusion matrices for 180 signals (60 per catalog) at four different SNR values.}
\label{matrices}
\end{figure}

Using this procedure over four different SNR values we obtain the classification results summarized in the confusion matrices of Figure \ref{matrices}. The classification of {\tt Dim} and {\tt SG} signals is entirely successful for SNR=20 and 15, while around 85\% {\tt Mur} signals are correctly classified. The results worsen significantly for SNR=10, with {\tt Dim} signals remaining the best identified ones, at 75\%. At SNR=5, the algorithm tends to classify a majority of signals as {\tt Mur} irrespective of their actual class (most hits are in the horizontal middle row in the bottom-right panel of Fig.~\ref{matrices}). While this is an undesirable result it is, however, not surprising as our method does not incorporate any criterion to classify a signal as noise and by their stochastic nature {\tt Mur} signals are the most difficult ones to separate from noise. A possible reason for the misclassification could be a failed denoising step in which the algorithm is unable to identify anything similar to the original signal. Another reason may be the generation of spurious signals from noise during the denoising step, as illustrated in Section 4.2. Such spurious signals could be classified as {\tt Mur} if matched closely those of the catalog. While higher values of $\boldsymbol \lambda^{\rm denoise}$ could avoid generating those, the tradeoff would be for the denoising to miss actual GW signals in low SNR conditions.

In an attempt to improve the performance of our algorithm when it comes to separating {\tt Mur} signals from noise, we run 60 different realizations of pure noise ``signals'' through our algorithm. We then use the resulting pure noise reconstructions to develop a way to classify signals as noise. By calculating the mean value of the WD of the pure noise classifications and subtracting the mean standard deviation, we set maximum values for WD$_{\rm Dim}$, WD$_{\rm Mur}$ and WD$_{\rm SG}$ over which a signal would be classified as noise. We find that this method still correctly classifies most {\tt Dim} and {\tt SG} signals at SNR=20 and SNR=15 without dismissing them as noise, while at SNR=10, 27\% of {\tt Dim} and 50\% of {\tt SG} signals are dismissed as noise. However, out of the remaining signals, 93\% of {\tt Dim} and 84\% of {\tt SG} signals are correctly identified. For SNR=5, around 70\% of both types of signals are dismissed as noise, while 50\% of the remaining classifications are successful. However, with this new method, around 60-80\% of Mur signals are classified as noise irrespective of the SNR. Again, these results indicate that distinguishing {\tt Mur} signals from pure noise represents a challenge to our algorithm.

\subsection{Comparison with SMEE}

It is worth comparing our findings with those obtained by SMEE~\citep{Logue:2012,Powell:2016,Powell:2017,Powell:2018,roma2019astrophysics}. As mentioned in the introduction SMEE classifies CCSN signals using principal component analysis and Bayesian model selection. To make a fair comparison, we contrast our results with those of SMEE in the case of one LIGO detector only, rather than with those of a three-detector network. 

SMEE correctly identifies signals that are already included in their model, namely 100\% of the magneto-rotational signals and around 95\% of the neutrino signals~\citep{smee}. If the signals are not included in the model, over 80\% of the neutrino signals and over 95\% of the magneto-rotational signals are correctly identified by SMEE. These numbers are comparable to our best results for SNR=20 and 15.

We note that besides the different methodological approach to reconstruct and classify signals between SMEE and our work, there are other aspects to consider that are relevant in the comparison. First, the majority of magneto-rotational signals classified by SMEE correspond to SNR values in the range 20-70, with some reaching up to 140, i.e.~much higher values than ours. Most of the neutrino signals, however, have SNR of 5 to 20, comparable to our values, with only a small set of signals reaching up to 30. Furthermore, \citet{smee} incorporate a criterion based on Bayesian analysis to identify an incoming signal as pure noise, something which is missing in our approach. The percentages of correctly classified signals mentioned above for SMEE take this into account and they correspond to signals identified as GW signals (those identified as noise were discarded). For signals included in the model, 73\% of neutrino signals and 9\% of magneto-rotational signals were identified as noise by SMEE. For those not included in the model, it amounted to 80\% and 32\%, respectively. 

From this comparison we conclude that learned dictionaries yield satisfactory results when identifying CCSN signals at high-enough SNR (15-20) and behave as expected when lowering the SNR.

\section{Conclusion}

In this paper we have presented an algorithm based on learned dictionaries to reconstruct and classify core-collapse supernova signals according to their morphology. CCSN are a prime source of gravitational waves with frequencies that lie within the sensitivity band of the current network of advanced, ground-based detectors~\citep{Abdikamalov:2020}. The analysis of those waveforms could potentially reveal the underlying explosion mechanism as has been shown with the so-called Supernova Model Evidence Extractor in a number of investigations~\citep{Logue:2012,Powell:2016,Powell:2017,Powell:2018,roma2019astrophysics}, a method based on principal component analysis and Bayesian model selection. Those works motivated us to find out whether the results of SMEE could be reproduced with a supervised dictionary-learning algorithm as a suitable complementary approach.

To carry out this comparison we have used the same waveform signals employed by SMEE, namely waveforms from two publicly available catalogs obtained from numerical simulations of neutrino-driven CCSN explosions~\citep{Murphy:2009} as well as magneto-rotational explosions~\citep{Dimmelmeier:2008}. In addition, we have used a third `mock' catalog of simulated sine-Gaussian waveforms not only to increase the types of signals in our study but also to enlarge the complexity for the algorithm. All signals have been injected into coloured Gaussian noise to simulate the background noise of Advanced LIGO in its broadband configuration and scaled to a freely-specifiable SNR. By using a quantitative criterion based on the Wasserstein distance we have constructed confusion matrices at four different SNR. Those matrices have shown that the behavior of the algorithm at SNR=20 and 15 is fairly satisfactory, with nearly all {\tt Dim} and {\tt SG} signals being correctly identified, while only around 15\% of the {\tt Mur} signals being misclassified.

It is not surprising that {\tt Mur} signals are harder to classify than the rest. They intrinsically resemble noise (characterized by a stochastic time series), they involve a large range of frequencies, their shapes are less generic, the signals are longer, and, perhaps most importantly, the amount of signals in the catalog is quite small (only 11) which impacts the training and classification. We note that the algorithm could be improved by devising a criterion for when a signal can be classified as `Noise'. In our current setup and at our lowest SNR there is a tendency for around 70\% of signals to be classified as {\tt Mur}. Discarding those signals which are identified as noise could possibly improve the success rate of our classifications. This is of particular interest after comparing our results with those of~\cite{smee}, which do include a `Noise' category. Using a one-detector-only setup and signals not included in the model, which is comparable to our own setup, SMEE classified 32\% magneto-rotational signals as pure noise. As most of those magneto-rotational signals have SNR$\ge20$, this is an encouraging comparison for our method, which correctly identified all {\tt Dim} signals at SNR=15-20. 

A drawback of our algorithm is the appearance of spurious signals when working in pure-noise conditions, which could lead to both misclassification of legitimate signals and false positives. While in an ideal scenario a LASSO regression would report no signal when applied to an only-noise time series, in our case it does return a signal, and this depends on the value of the regularization parameter. While working with higher values of $ \lambda^{\rm denoise}$ might reduce the number of false signals, it could also carry the tradeoff of increasing false negatives, particularly for more noisy scenarios. Furthermore, the classification of CCSN signals relies on having access to simulated waveforms such as the {\tt Dim} and {\tt Mur} catalogs employed in this work, which are computationally costly to obtain. The scarcity of available numerical waveforms (i.e.~data to train the algorithm) is an obvious limiting factor of our method. Despite possible future improvements, our current results show it is possible to use LASSO regression to extract gravitational-wave signals injected in Gaussian noise using trained dictionaries, and to classify them depending on their morphology after being reconstructed out of pre-existing waveforms. 

On a final note, the~\citet{Dimmelmeier:2008} and \citet{Murphy:2009} waveforms employed in this study, while useful for the comparison with SMEE, would not be used to determine the explosion mechanism in an actual case of a supernova detection candidate. Among the reasons to discard them is the fact that the~\citet{Dimmelmeier:2008} waveforms are both  axisymmetric and fairly short, since they only include the early bounce part of the signal, while the~\citet{Murphy:2009} waveforms are also only 2D and there are currently more accurate 3D neutrino-driven waveforms available. Performing a further comparison study using a larger set of improved CCSN waveforms is an interesting idea to pursue as a follow-up of this work,

\section*{Acknowledgements}
We thank Jade Powell for her careful reading of the manuscript and useful comments. This work has been supported by the Spanish Agencia Estatal de Investigaci\'on (grant PGC2018-095984-B-I00) and by the Generalitat Valenciana (grant PROMETEO/2019/071). 



\bibliographystyle{mnras}
\bibliography{./referencias}



\appendix

\section{Optimization of the dictionary parameters}

This appendix summarizes the procedures through which we determine the best parameters of the dictionaries. First, we choose an optimal $\lambda^{\rm train}$ for each of the dictionaries created. We qualitatively choose lower and upper limits to generate each dictionary following the criteria detailed in Section 4.1. Then, we train four \textbf{D}$^{\rm train}$ dictionaries for each $n$, $p$, and signal type, with four different values of $\lambda^{\rm train}$. Finally, we use all the dictionaries to denoise signals of their corresponding type injected in noise with SNR=20, and we compute the WD from each denoised signal \textbf{y} to the original signal \textbf{u}. The optimal values of $\lambda^{\rm train}$ for which the smallest WD are found are reported in Table~\ref{lambdatrainopt}.

\begin{table}
\centering
\caption{Optimal values of the regularization coefficients $\lambda^{\rm train}$ for dictionaries trained at different $n$ and $p$. The Wasserstein distances shown are the mean from all denoisings of their corresponding signal type with SNR=20. The left and right values shown in the last two columns indicate the results for $p=100$ and $p=200$, respectively.}
\begin{tabular}{c|ccccc}
\hline \hline
\textbf{\begin{tabular}[c]{@{}c@{}}Signal\\ type\end{tabular}} & \textbf{n} & \textbf{$\boldsymbol\lambda^{\rm train}_{\rm min}$} & \textbf{$\boldsymbol\lambda^{\rm train}_{\rm max}$} & \textbf{\begin{tabular}[c]{@{}c@{}}$\boldsymbol\lambda^{\rm train}_{\rm opt}$\\ 100$|$200\end{tabular}} & \textbf{\begin{tabular}[c]{@{}c@{}}WD$_{\rm opt}\times 10^{-2}$\\ 100$|$200\end{tabular}} \\ \hline
      & 64     & 0.2        & 0.8        & 0.4$|$0.2                                  & 2.08$|$1.85                   \\
Dim     & 128    & 0.3        & 1.2        & 1.2$|$1.2                                 & 2.28$|$2.04                    \\
      & 256    & 0.6        & 1.5        & 1.2$|$1.5                                 & 2.71$|$2.25           \\
      & 512    & 0.6        & 2.0         & 2.0$|$1.5                                  & 3.09$|$2.56          \\ \hline
      & 64     & 0.2        & 0.8        & 0.2$|$0.2                                 & 1.38$|$0.86            \\
Mur     & 128    & 0.4        & 0.9        & 0.4$|$0.6                                 & 2.01$|$1.62     \\
      & 256    & 0.4        & 1.1        & 1.1$|$0.4                                 & 3.25$|$2.54        \\
      & 512    & 0.6        & 1.7        & 1.0$|$1.4                                  & 4.05$|$3.67       \\ \hline
      & 128    & 0.2        & 0.3        & 0.25$|$0.30                                 & 9.55$|$10.4                                   \\
SG   & 256    & 0.55        & 0.7        & 0.70$|$0.65                                 & 7.42$|$10.4   \\
      & 512    & 0.7        & 0.85        & 0.80$|$0.85                                 & 6.49$|$8.95                                \\
      & 1024    & 1.2        & 1.9        & 1.20$|$1.70                                 & 4.87$|$4.11                               \\ \hline \hline
\end{tabular}
\label{lambdatrainopt}
\end{table}

We note that the WD are not computed over the entire duration of the signals but rather over three specific ranges for each signal type. 
Those are chosen to avoid the optimization process from evaluating signals containing mostly or entirely zeros. The specific ranges are 0.71-0.84~s for {\tt Dim} signals, 0.15-1.29~s for {\tt Mur} signals, and 0.79-1.19~s for {\tt SG} signals. These are the ranges for which we optimize both $\lambda^{\rm train}$ and $\lambda^{\rm denoise}$.

\begin{figure}
\centering
\includegraphics[scale=0.21]{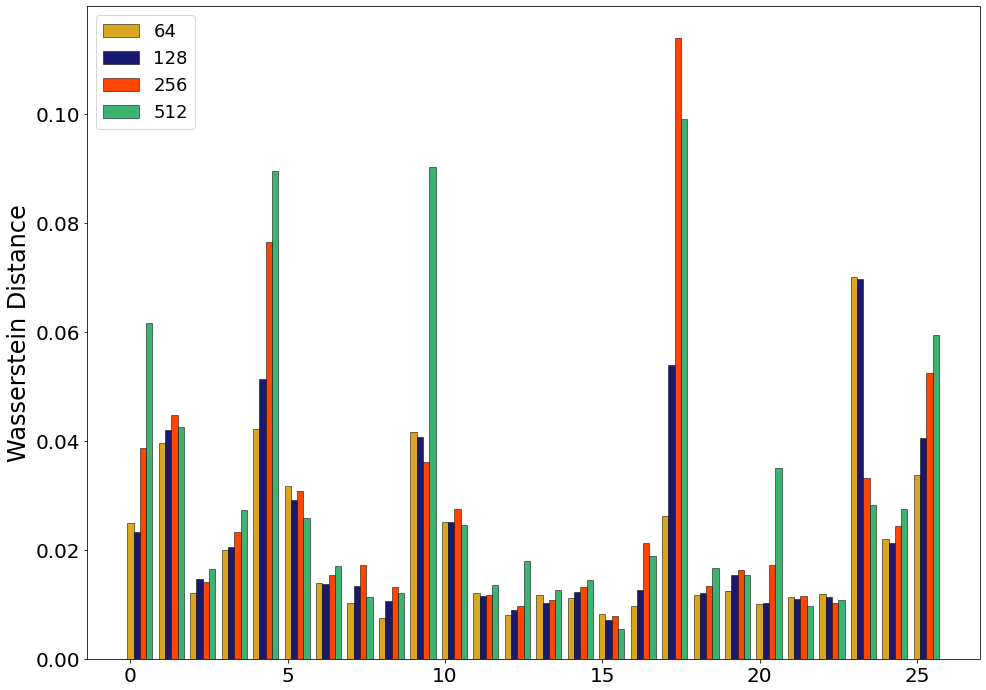} 
\includegraphics[scale=0.21]{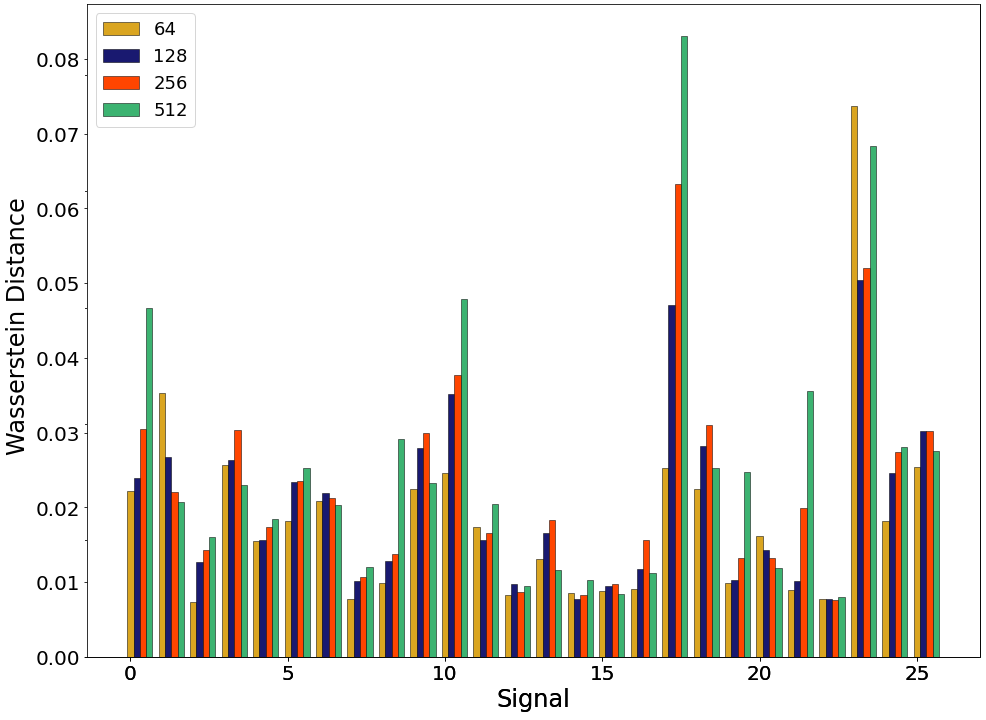}
\caption{
Histograms of the WD for the 26 {\tt Dim} validation signals injected in noise at SNR=20, denoised with different dictionaries as indicated by the legends. The top and bottom panels show the results for $p = 100$ and $p = 200$, respectively.}
\label{dimdicts}
\end{figure}

Once the optimal $\lambda^{\rm train}_{\rm opt}$ is chosen for each dictionary we have to decide the optimal values of $n$ and $p$. For this we carry out both a quantitative analysis through the WD and a qualitative one to check the results. We begin by comparing the WD calculated for every signal with each dictionary. We start with the {\tt Dim} dictionaries and signals, displaying the results of the reconstructions in Figure \ref{dimdicts}. We see that there is no unique value of $n$ that yields the best reconstruction for every single signal. From this plot it appears that $n=64$, $p=200$ gives on average the best results, as can be seen on Table \ref{lambdatrainopt} as well. In general larger values of $p$ will result in better reconstructions, though at a higher computational cost. 

However, a qualitative analysis shows that $n=64$ dictionaries build more oscillatory signals, as the shorter length of the atoms results in the introduction of larger amounts of noise. Larger atom lengths could reduce this at the cost of missing smaller (physical) oscillations in the signals. Therefore, despite the WD computes the $n=64$ signals as the best scenario, minimizing the amount of noise in the denoising process is preferable for the subsequent reconstructions. As our aim is to develop an algorithm that classifies signals by their type, we thus select a dictionary by requiring it to be able to identify the overall shape of a given signal without introducing too much noise that might interfere with the classification process. With these considerations in mind we decide on the $n=256$, $p=200$ dictionary. 

\begin{figure}
\centering
\includegraphics[scale=0.21]{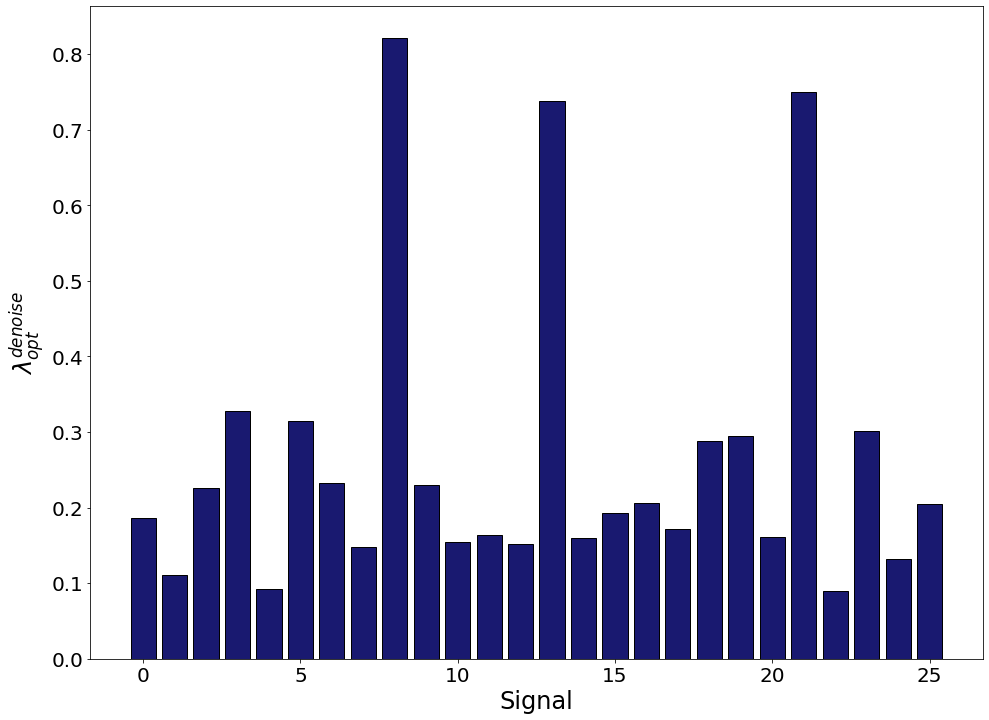} 
\caption{Histogram of the optimal values of the regularization parameter $\lambda^{\rm denoise}_{\rm opt}$ for the {\tt Dim} signals. The values correspond to SNR=20 and to a \textbf{D}$^{\rm train}_{\rm Dim}$ dictionary with $n=256$ and $p=200$.}
\label{lambdadim}
\end{figure}

The next parameter that must be evaluated is $\lambda^{\rm denoise}$. 
Up to this point in the algorithm for each denoising we have optimized every signal and dictionary so that it would find the value of $\lambda^{\rm denoise}$ that minimizes the WD. Moving forward, we want to find a fixed value $\lambda^{\rm denoise}_{\rm opt}$ which will be used to denoise every signal (one per catalog) as, in a realistic setup, we do not have access to the original signal $\textbf{u}$ to optimize the WD. Figure \ref{lambdadim} shows $\lambda^{\rm denoise}_{\rm opt}$ for the 26 {\tt Dim} signals. The values are bounded between around 0.1 and 0.8, with very few of them going above 0.3. Therefore, we discard the three highest values and choose the mean value of $\lambda^{\rm denoise}_{\rm Dim}=0.2$, which we expect will produce good reconstructions for most {\tt Dim} signals. 

\begin{figure}
\centering
\includegraphics[scale=0.21]{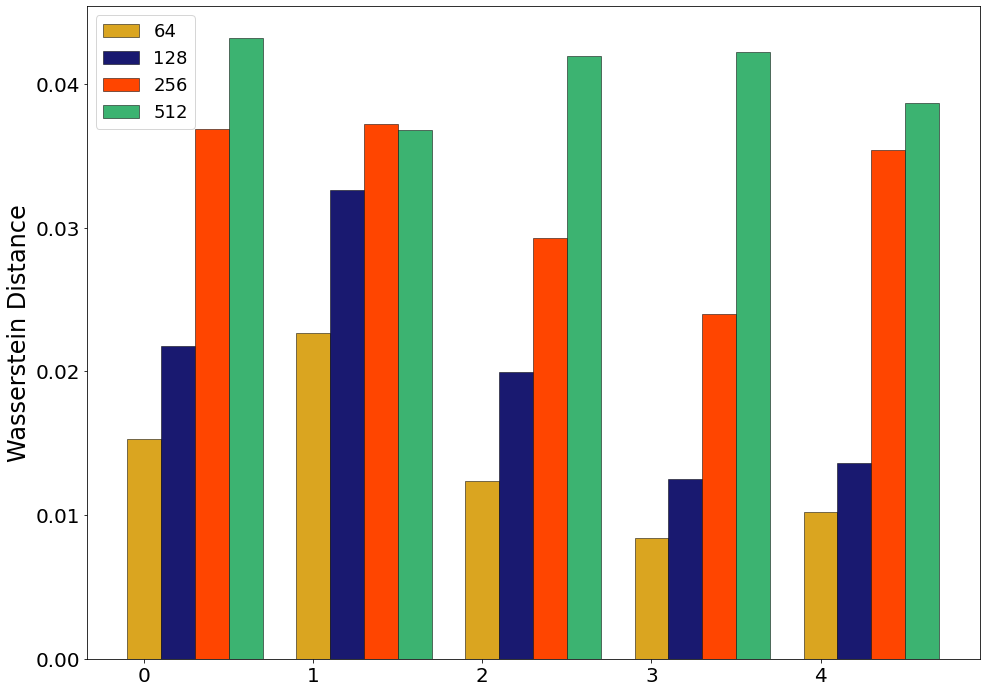} 
\includegraphics[scale=0.21]{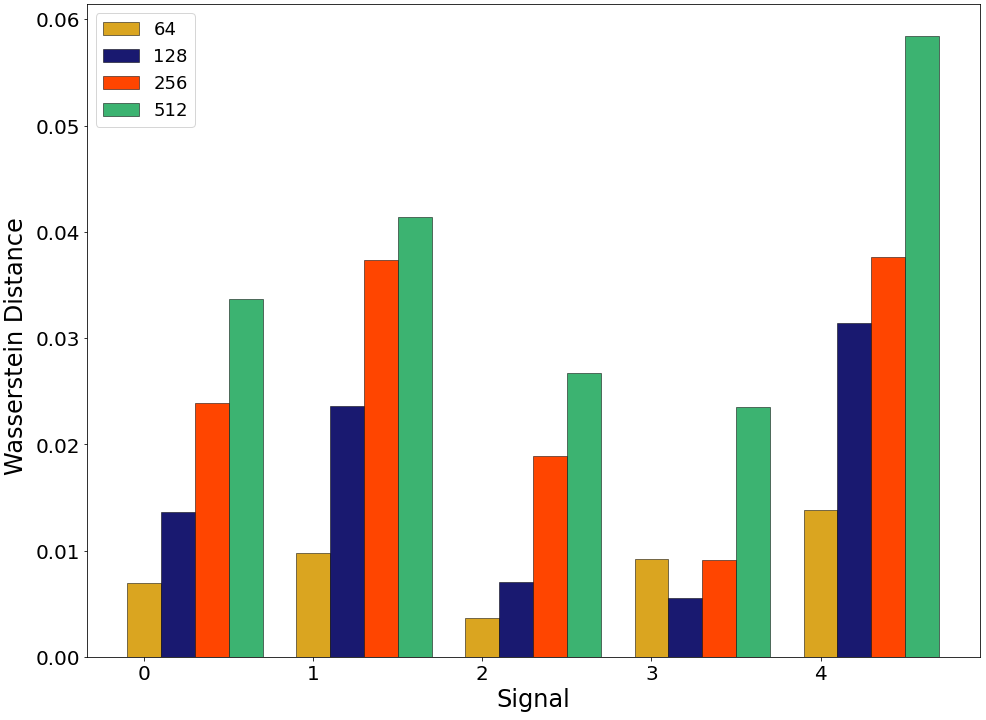} 
\caption{
Histograms of the WD for the 5 {\tt Mur} validation signals injected in noise at SNR=20, denoised with different dictionaries as indicated by the legends. The top and bottom panels show the results for $p = 100$ and $p = 200$, respectively.}
\label{murdicts}
\end{figure}

\begin{figure}
\centering
\includegraphics[scale=0.21]{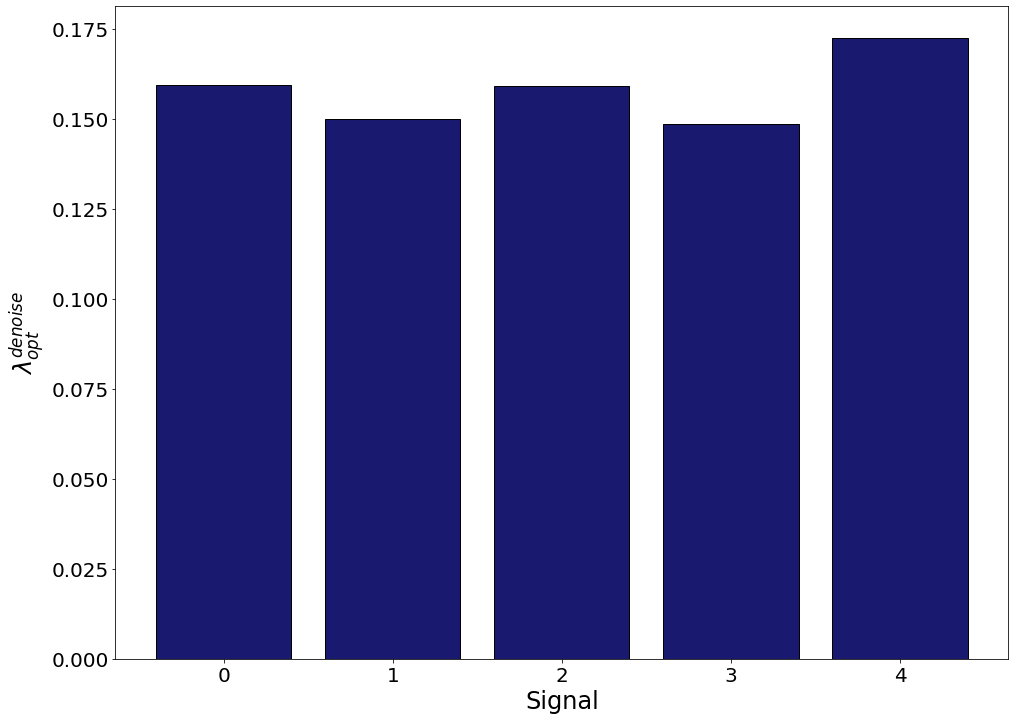} 
\caption{
Histogram of the optimal values of the regularization parameter $\lambda^{\rm denoise}_{\rm opt}$ for the {\tt Mur} signals. The values correspond to SNR=20 and to a \textbf{D}$^{\rm train}_{\rm Mur}$ dictionary with $n=64$ and $p=200$.}
\label{lambdamur}
\end{figure}

The optimization process we have just described for the {\tt Dim} signals has to be repeated for the other types of signals, {\tt Mur} and {\tt SG}. 
Figure \ref{murdicts} displays the results for the Mur dictionaries. 
From this figure 
we can safely rule out the cases $n=256$ and $n=512$ as the corresponding values of the WD are too large to yield acceptable reconstructions. The tradeoff between $n=64$ and $n=128$, as discussed earlier, is a better reproduction of the oscillations for smaller atom sizes
against the presence of more noise throughout the signal. Figure \ref{murdicts} shows that the $n=64$, $p=200$ dictionary is the best choice for practically all signals. We therefore choose this dictionary as the results can be more easily identified with {\tt Mur} signals despite the extra noise. In general, we expect the reconstructions of {\tt Mur} signals to yield the worst results of all signal types considered. In addition, Figure \ref{lambdamur} depicts the histogram of the optimal values of $\lambda^{\rm denoise}_{\rm opt}$ for the {\tt Mur} signals. We choose $\lambda^{\rm denoise}_{\rm Mur}=0.16$, as the values are bounded between 0.15 and 0.175.

\begin{figure}
\centering
\includegraphics[scale=0.21]{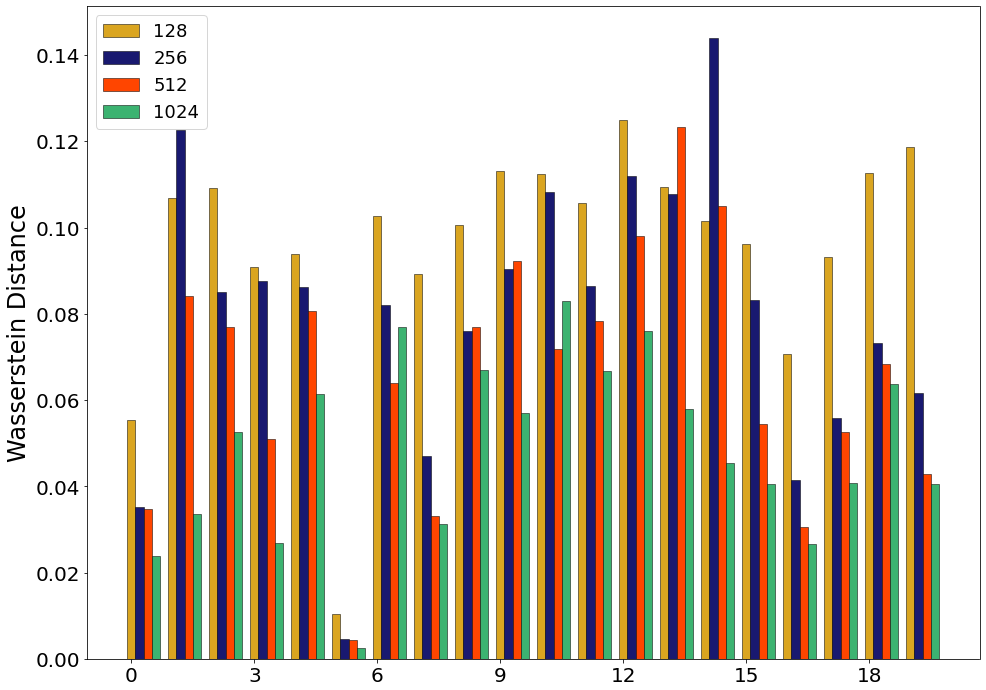} 
\includegraphics[scale=0.21]{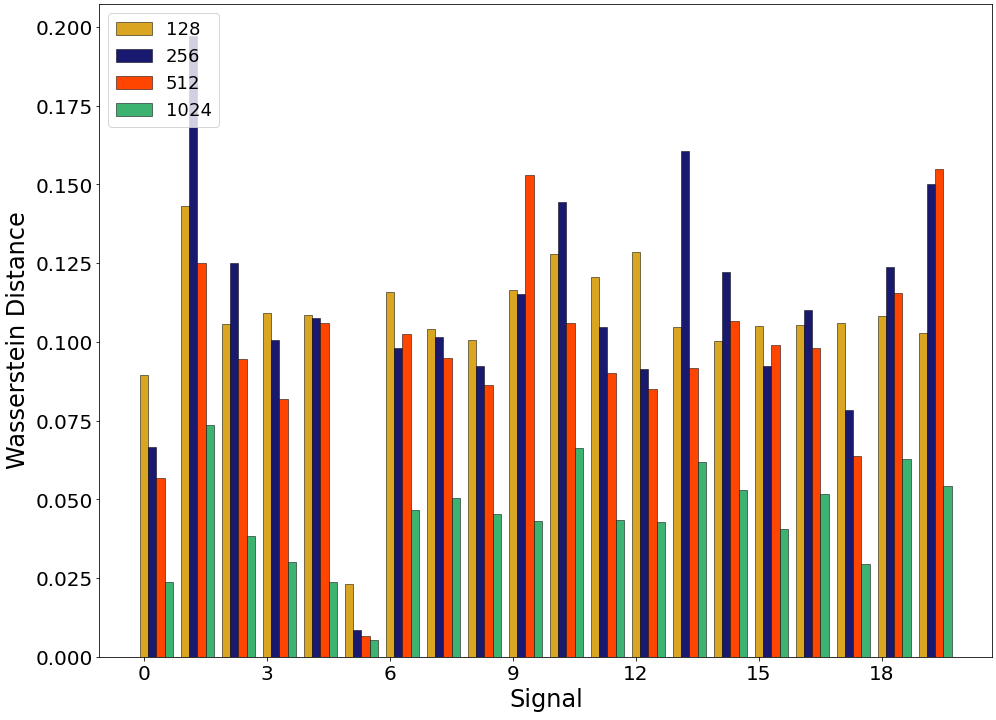} 
\caption{
Histograms of the WD for the half the {\tt SG} validation signals injected in noise at SNR=20, denoised with different dictionaries as indicated by the legends. The top and bottom panels show the results for $p = 100$ and $p = 200$, respectively.}
\label{glitchdicts}
\end{figure}

\begin{figure}
\centering
\includegraphics[scale=0.21]{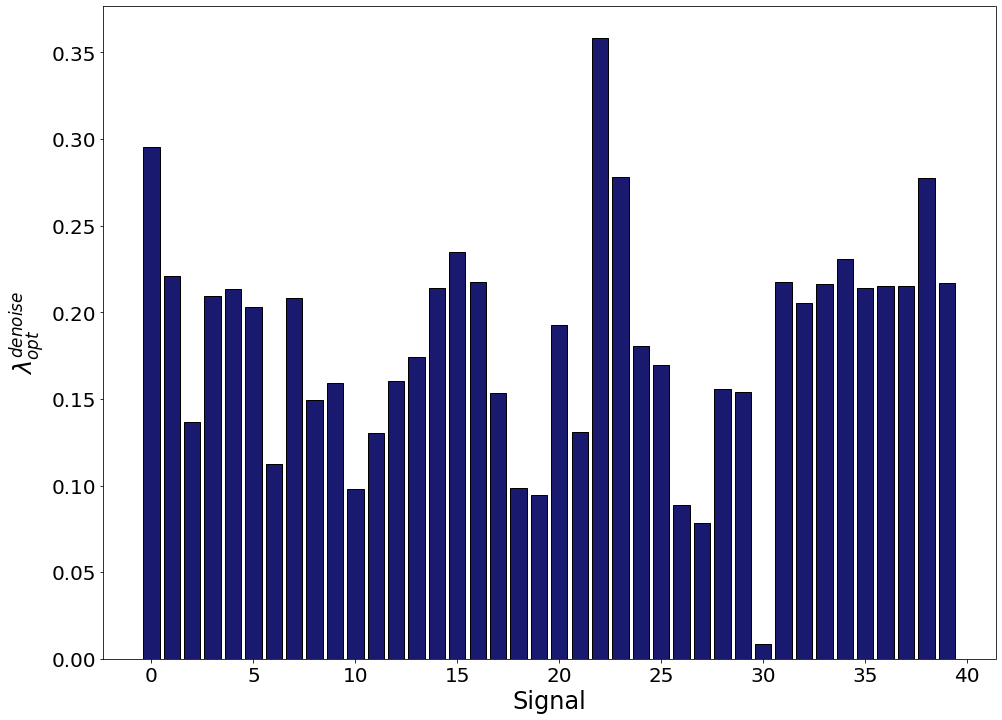} 
\caption{
Histogram of the optimal values of the regularization parameter $\lambda^{\rm denoise}_{\rm opt}$ for the {\tt SG} signals. The values correspond to SNR=20 and to a \textbf{D}$^{\rm train}_{\rm Dim}$ dictionary with $n=1024$ and $p=200$.}
\label{lambdaglitch}
\end{figure}

Finally, the estimation of the parameters for the {\tt SG} signals is displayed in Figure \ref{glitchdicts} and Figure \ref{lambdaglitch}. For clarity we only display in Fig.~\ref{glitchdicts} the results for half of the validation signals (i.e.~only 20 {\tt SG} signals). Nevertheless, the conclusion that $n=1024$ and $p=200$ produces the best reconstructions applies to the entire set. On the other hand, Figure \ref{lambdaglitch} reveals that the values of $\lambda^{\rm denoise}_{\rm opt}$ show larger variations for {\tt SG} signals than for the previous two sets of signals. As a result we choose the mean value of the entire sample, namely $\lambda^{\rm denoise}_{\rm SG}=0.17$.

As a summary Table \ref{dictionaryparameters} reports the denoising parameters we use in the signal classification process, namely the values of $n$ and $\lambda^{\rm train}$ to train our three $\textbf{D}^{\rm train}$ dictionaries and the corresponding $\lambda^{\rm denoise}$ values for the LASSO regression. Although finding the best value of $\lambda^{\rm denoise}$ for each separate SNR would yield better results, and even more so optimizing $\lambda^{\rm denoise}$ for each individual signal, our method has been developed to be applicable to CCSN signals of unknown parameters, including its SNR.

\begin{table}
\centering
\caption{Optimal values of the parameters of the algorithm inferred from the denoising process for each signal type. All dictionaries used will have $p=200$.}
\begin{tabular}{c|ccc}
\hline \hline
\textbf{Type} & \textbf{n} & \textbf{$\boldsymbol\lambda^{\rm train}$} & \textbf{$\boldsymbol\lambda^{\rm denoise}$} \\ \hline
Dim      & 256     & 1.5         & 0.2 \\
Mur      & 64     & 0.2         & 0.16         \\
SG    & 1024    & 1.7         & 0.17         \\ \hline \hline
\end{tabular}
\label{dictionaryparameters}
\end{table}

\bsp	
\label{lastpage}
\end{document}